\documentclass[11pt]{article}

\newcommand{\blind}{0}

\newcommand{\coloreditsfin}{0} 

\newcommand{\reditfin}[1]{\if1\coloreditsfin\textcolor{red}{\st{#1}}\fi}
\newcommand{\beditfin}[1]{\if1\coloreditsfin\textcolor{blue}{#1}\fi\if0\coloreditsfin\textcolor{black}{#1}\fi}

\usepackage{graphicx,amsthm,amssymb,multirow,amsmath,soul,mathtools}

\usepackage{etoolbox,refcount} 
\usepackage{thmtools}
\usepackage{environ}
\RequirePackage{etex}

\makeatletter 

\renewcommand\section{\@startsection {section}{1}{\z@}%
                                   {-3.5ex \@plus -1ex \@minus -.2ex}%
                                   {2.3ex \@plus.2ex}%
                                   {\normalfont\fontfamily{phv}\fontsize{16}{19}\bfseries}}
\renewcommand\subsection{\@startsection{subsection}{2}{\z@}%
    {-3.25ex\@plus -1ex \@minus -.2ex}%
    {1.5ex \@plus .2ex}%
   {\normalfont\fontfamily{phv}\fontsize{14}{17}\bfseries}}
\renewcommand\subsubsection{\@startsection{subsubsection}{3}{\z@}%
    {-3.25ex\@plus -1ex \@minus -.2ex}%
   {1.5ex \@plus .2ex}%
   {\normalfont\normalsize\fontfamily{phv}\fontsize{14}{17}\selectfont}}

\providecommand{\@fourthoffour}[4]{#4}
\newcommand\fixstatement[2][\proofname\space of]{%
  \ifcsname thmt@original@#2\endcsname
    \AtEndEnvironment{#2}{%
      \xdef\pat@label{\expandafter\expandafter\expandafter
        \@fourthoffour\csname thmt@original@#2\endcsname\space\@currentlabel}%
      \xdef\pat@proofof{\@nameuse{pat@proofof@#2}}%
    }%
  \else
    \AtEndEnvironment{#2}{%
      \xdef\pat@label{\expandafter\expandafter\expandafter
        \@fourthoffour\csname #1\endcsname\space\@currentlabel}%
      \xdef\pat@proofof{\@nameuse{pat@proofof@#2}}%
    }%
  \fi
  \@namedef{pat@proofof@#2}{#1}%
}
\globtoksblk\prooftoks{1000}
\newcounter{proofcount}

\usepackage{hyperref} 
\hypersetup{ 
    colorlinks=true,
    linkcolor=black,
    filecolor=black,
    citecolor=black,
    urlcolor=blue}

\usepackage{url}
\usepackage{thm-restate}
\usepackage{subfig} 
\usepackage{wrapfig}
\usepackage[ansinew]{inputenc}
\usepackage{tikz}
\usepackage[colorinlistoftodos,prependcaption,textsize=tiny]{todonotes} 
\usepackage{xargs}
\newcommandx{\QEug}[2][1=]{\todo[linecolor=blue,backgroundcolor=blue!25,bordercolor=blue,#1]{#2}}
\usepackage{setspace}
\usepackage{textcomp}
\usepackage{enumitem}
\usepackage{enumerate}
\usepackage{multirow}       
\usepackage{makecell}       
\setcellgapes{4pt}          
\usepackage{booktabs}       
\usepackage{array}
\newcolumntype{P}[1]{>{\centering\arraybackslash}p{#1}}
\usepackage{tabulary}
\usepackage{caption}        
\usepackage{nccmath} 
\usepackage{mathrsfs} 
\usetikzlibrary{fit,positioning}
\usepackage{setspace}
\usepackage[font=it,labelfont=bf]{caption} 
\usepackage{bbm} 
\usepackage{csquotes} 

\usepackage[american]{babel}
\usepackage[backend=biber, style=authoryear,
defernumbers=true,
uniquename=false,
uniquelist=false,
citestyle=authoryear,
sorting=anyvt,maxcitenames=3,
maxbibnames=25,
giveninits=true, 
natbib]{biblatex}

\renewbibmacro*{volume+number+eid}{
  \printfield{volume}%
  \printfield{number}%
  \setunit{\addcomma\space}%
  \printfield{eid}}
\DeclareFieldFormat[article]{number}{\mkbibparens{#1}}
\DeclareFieldFormat[article]{volume}{\textbf{#1}}
\DeclareFieldFormat{pages}{#1}
\DeclareFieldFormat[article]{title}{#1} 
\renewbibmacro{in:}{
  \ifentrytype{article}
    {}
    {\bibstring{in}%
     \printunit{\intitlepunct}}
}
     
\AtEveryBibitem{ 
\ifentrytype{article}{
    \clearfield{eprint}%
    \clearfield{url}
    \clearfield{doi}
}{}
\ifentrytype{book}{\clearfield{url}
    \clearfield{doi}
}{}
}     
     
\usepackage{wrapfig} 

\usepackage{cleveref}
\theoremstyle{plain}
\newtheorem{thrm}{Theorem}

\newtheoremstyle{indent}
  {}
  {}
  {\itshape}
  {\parindent}
  {\bfseries}
  {.}
  {.5em}
  {}
\theoremstyle{indent}

\usepackage[margin=1in]{geometry}

\usepackage{array}
\newcolumntype{L}[1]{>{\raggedright\let\newline\\\arraybackslash\hspace{0pt}}m{#1}}
\newcolumntype{C}[1]{>{\centering\let\newline\\\arraybackslash\hspace{0pt}}m{#1}}
\newcolumntype{R}[1]{>{\raggedleft\let\newline\\\arraybackslash\hspace{0pt}}m{#1}}

\usepackage{nicefrac,xfrac}

\newcommand{\bm}[1]{\boldsymbol{#1}} 
\newcommand{\importerSet}{\mathcal{B}}
\newcommand{\outletSet}{\mathcal{A}}
\newcommand{\numOutlets}{|\outletSet|}
\newcommand{\numImporters}{|\importerSet|}

\newcommand{\importerSFPrates}{\theta}
\newcommand{\outletSFPrates}{\eta}

\newcommand{\consolSFPfunc}{z^\star}
\newcommand{\consolInaccFunc}{z}

\newcommand{\transMat}{\bm{Q}}
\newcommand{\diagSens}{s}
\newcommand{\diagSpec}{r}

\newcommand{\numTests}{n}
\newcommand{\testResultRV}{Y}
\newcommand{\testResultpoint}{y}
\newcommand{\importerLabelRV}{B}
\newcommand{\importerLabelpoint}{b}
\newcommand{\outletLabelRV}{A}
\newcommand{\outletLabelpoint}{a}
\newcommand{\SFPrateSet}{\outletSFPrates,\importerSFPrates}
\newcommand{\SFPrateSetPrime}{\outletSFPrates',\importerSFPrates'}
\newcommand{\outletExpit}{\alpha}
\newcommand{\importerExpit}{\beta}
\newcommand{\expitConstantMean}{\gamma}
\newcommand{\expitConstantVar}{\nu}
\newcommand{\logitfunc}{g}
\newcommand{\expitfunc}{\logitfunc^{-1}}
\newcommand{\suchthat}{\mathrel{\mathop\supset}\kern-4.0pt$-$\kern-1.0pt$-~$}

\newcommand{\priorHessConst}{C_2}

\newcommand{\coupleSet}{\mathcal{E}}

\newcommand{\dataSet}{\bm{d}}
\newcommand{\priorFunc}{p}
\newcommand{\postFunc}{p}
\newcommand{\elemSet}{\mathcal{K}}
\newcommand{\elem}{k}

\newcommand{\testPosAvg}{\Bar{\consolInaccFunc}}

\usepackage[markup=underlined]{changes}

\usepackage{changepage}

\usepackage[title]{appendix}

\long\def\proofatend#1\endproofatend{%
  \edef\next{\noexpand\begin{proof}[Proof of \pat@label]}%
  \toks\numexpr\prooftoks+\value{proofcount}\relax=\expandafter{\next#1\end{proof}}
  \stepcounter{proofcount}}

\def\printproofs{%
  \count@=\z@
  \loop
    \the\toks\numexpr\prooftoks+\count@\relax
    \ifnum\count@<\value{proofcount}%
    \advance\count@\@ne
  \repeat}

\makeatother

\fixstatement[Theorem]{thrm}
\fixstatement[Lemma]{lemma}
\fixstatement[Proposition]{prop}
\fixstatement[Corollary]{corol}
\fixstatement[Definition]{dfn}

\def\rot{\rotatebox}
\usepackage{rotating} 

\usepackage{xcolor,cancel}

\usepackage{geometry}
\usepackage{pdflscape}

\usepackage[linesnumbered,ruled,vlined]{algorithm2e}

\SetCommentSty{mycommfont}
\SetArgSty{textnormal}

\SetKwInput{KwInput}{Input}                
\SetKwInput{KwOutput}{Output}              

\usepackage[export]{adjustbox}

\begin{document}
\def\spacingset#1{\renewcommand{\baselinestretch}%
			{#1}\small\normalsize} \spacingset{1}
	\if0\blind
		{ \title{\bf Inferring sources of substandard and falsified products in pharmaceutical supply chains}
			\author{Eugene Wickett $^a$, Matthew Plumlee $^a$, Karen Smilowitz $^a$, \\ Souly Phanouvong $^b$, and Victor Pribluda $^c$ \\
			$^a$ Industrial Engineering and Management Sciences, Northwestern University, \\ Evanston, Illinois 
			\\ $^b$ Promoting the Quality of Medicines Plus (PQM+) Program, \\ U.S. Pharmacopeial Convention (USP), Rockville, Maryland \\
			$^c$  U.S. Pharmacopeial Convention (USP), Rockville, Maryland
			}
			\date{}
			\maketitle
		} \fi
	\if1\blind
	{\title{\bf Inferring sources of substandard and falsified products in pharmaceutical supply chains}
	\author{Author information is purposely removed for double-blind review}
		    \date{}
		    \maketitle
		} \fi

\begin{abstract}
Substandard and falsified pharmaceuticals, prevalent in low- and middle-income countries, substantially increase levels of morbidity, mortality and drug resistance.
Regulatory agencies combat this problem using post-market surveillance by collecting and testing samples where consumers purchase products.
Existing analysis tools for post-market surveillance data focus
attention on the locations of positive samples.
This paper looks to expand such analysis through underutilized supply-chain information to provide inference on sources of substandard and falsified products.
We first establish the presence of unidentifiability issues when integrating this supply-chain information with surveillance data.
We then develop a Bayesian methodology for evaluating substandard and falsified sources that extracts utility from supply-chain information and mitigates unidentifiability while accounting for multiple sources of uncertainty.
Using de-identified surveillance data, we show the proposed methodology to be effective in providing valuable inference.
\end{abstract}

\noindent%

\spacingset{1.5} 

\section{Introduction} \label{sec:Introduction}
Substandard and falsified pharmaceuticals (SFPs) are a pressing global health issue.
Recent studies estimate that around ten percent of medical products in low-and middle-income countries are unsuitable for consumption; estimates indicate higher burdens depending on the disease or assessment methodology \parencite{who2018,ozawa2018,koczwara2017}.
Mortality estimates assert that SFPs lead to 450,000 preventable deaths every year \parencite{karunamoorthi2014}.
SFPs also contribute to the growing worldwide threat of drug resistance \parencite{who2017}, as well as diminished public confidence in health systems \parencite{cockburn2005}.

\vspace{-16pt}
\subsection{Post-market surveillance}
Medical products regulators ensure pharmaceutical quality through different activities conducted throughout the manufacturing and distribution processes.
Following data monitored at
\if0\blind
United States Pharmacopeia,
\fi
\if1\blind
[\textit{redacted organization engaged in medical products regulation}],
\fi
this paper considers post-market surveillance (PMS) where regulators collect samples from consumer-facing outlets and test those samples for compliance with registration specification \parencite{nkansah2018}.
The goal of PMS is estimation of SFP prevalence in regulatory domains and identification of sources of either substandard or falsified pharmaceuticals.
Usual PMS in low- and middle-income countries comprises of three stages.
The first stage selects a subset of locations that distribute pharmaceuticals to consumers, and the second stage collects and tests pharmaceuticals from these locations.
The third stage analyzes testing data and enforces corrective actions.
Corrective actions can include issuing warnings or recalls for particular brands or supply-chain locations.
Stretched regulatory budgets translate to limited PMS data: a single PMS activity may comprise a few hundred tests, used to evaluate an entire pharmaceutical indication, e.g., antimalarials.
Data constraints necessitate effective use of available metadata and regulatory domain knowledge to better understand SFP patterns.

Current methods for the analysis stage of PMS focus on establishing tolerance thresholds of SFP prevalence at sampled supply-chain locations.
Supply-chain information is regularly stored as part of PMS protocols.
The MEDQUARG guidelines of \citet{newton2009}, an industry standard for PMS, recommend collection of various supply-chain features of the outlet location and manufacturer of each sample.
The Medicines Quality Database (MQDB), featured in the case study of Section \ref{sec:casestudy}, captures PMS results submitted by dozens of participating national medical products regulators in line with the MEDQUARG guidelines \parencite{mqd2021}.
Each MQDB record contains testing results and associated supply-chain metadata such as manufacturer, manufacturer country, sampling location, and region of the sampling location.

Consideration of PMS within supply chains carries unique properties in the field of network detection.
SFP sources can be situated at any location from manufacturer to consumer; testing data from consumer-facing locations measure quality reflective of SFP sources throughout the supply chain beyond tested locations.
Thus, it is not clear whether a detected SFP is due to the consumer-facing location or an upstream supply-chain location.
Additionally, the supply-chain path of each sample is typically only partially known: labels are not applied every time a sample traverses the supply chain, and paths are only known probabilistically in some cases.
Different consumer-facing locations often share manufacturers or other upstream supply-chain locations.
Understanding these shared supply-chain connections can help regulators identify SFP sources.
In current practice, PMS data may be analyzed by manufacturer or aggregation of regional consumer-facing locations, but the information contained in supply-chain connections is underutilized.

\vspace{-10pt}
\subsection{Supply-chain PMS}
\vspace{-7pt}
\noindent
There is a need for PMS analysis methods that can infer the origin of SFP generation by modeling the paths of SFPs across separate supply-chain echelons.
An echelon is a collection of supply-chain locations that share a key attribute or function\reditfin{---for instance,}\beditfin{, such as} the collection of manufacturers or the collection of outlets that sell products.
SFP generation refers to either the degradation of product quality or the infiltration of falsified products.
The origin of an SFP is the location where an SFP is generated.
The origin can differ from where the SFP is detected.
For instance, pharmaceuticals can be produced according to good manufacturing practices but stored at a distribution warehouse where temperatures exceed allowable limits, causing degradation and resulting in substandard products.
Alternatively, an outlet can receive quality products, but sell a falsified substitute to the public while re-selling the quality products elsewhere.
This paper explores if identification of origins of SFP generation can be improved by incorporating supply-chain connections between consumer-facing testing locations and one upstream echelon.
We model only one additional upstream echelon due to PMS data availability common\reditfin{ in low-resource} \beditfin{to low- and middle-income} settings. 
While we model two echelons of a larger, more complex supply chain, this work is a step to expanding PMS capabilities through supply-chain information, even when such information is limited.

In our analysis of consumer-facing testing locations and an upstream echelon, we identify three types of uncertainty: fundamental unidentifiability, testing accuracy, and untracked supply-chain information.
Uncertainty due to fundamental unidentifiability results from only testing the lower echelon of a supply chain.
Confirmation of SFP generation at upstream locations is not possible without upstream testing; thus the aim is to examine if SFPs were generated at tested locations or further upstream, requiring additional investigation.
Uncertainty due to testing accuracy comes from imperfect testing equipment, human error, and inappropriate use of testing methods \parencite{kovacs2014}.
Testing accuracy is measured through sensitivity, which captures the ability to correctly detect SFPs, and specificity, which captures the ability to correctly detect quality products.
Uncertainty due to untracked supply-chain information arises when the path traversed by a sample is only known probabilistically.
Under untracked information, rather than knowing the exact supply-chain path a product takes to reach the sampled location, there is a known probability distribution for a sample's path across upper-echelon locations.
Our methodology accounts for these sources of uncertainty using a Bayesian framework that synthesizes testing data with available supply-chain information to infer SFP sources and thus guide regulator decisions.

\vspace{-10pt}
\subsection{Contributions}
\vspace{-5pt} 
\paragraph{Consideration of PMS in supply chains} This paper builds on existing PMS practice through incorporation of frequently available supply-chain information.
We use as an experiment the MQDB, which contains manufacturer labels as well as province and sub-region information for the consumer-facing location of each test.
Current practice does not synthesize PMS test results with supply-chain information towards inference of SFP sources.
Given that SFPs are recognized as a supply-chain problem---as described in Section \ref{sec:lit}---integrating readily available supply-chain information with testing results is a novel advance in PMS analysis.

\vspace{-10pt} 
\paragraph{Understanding unidentifiability} Whether SFP rates throughout a supply chain can be recovered through PMS has not been explored. 
By integrating testing data with supply-chain information, we establish unidentifiability of SFP rates in supply chains.
Establishing unidentifiability is a key contribution: we show SFP rates cannot be recovered through consideration of PMS testing results alone.
Understanding PMS results requires approaches that mitigate this unidentifiability.

\vspace{-10pt} 
\paragraph{General algorithms for\reditfin{ low-resource settings} \beditfin{low- and middle-income countries}}
Low- and middle-income countries require flexible analysis methods.
PMS data collection in these countries features considerable heterogeneity in available metadata.
PMS samples usually have a manufacturer label, and may also have a label designating one or more intermediate distributors.
The sampling location carries additional regional designations such as city or district.
Crucially, any of these designations may be critical to understanding SFP occurrence \parencite{pisani2019}.
Although frameworks like MEDQUARG for standardizing the collection of such metadata have been proposed, data collection from country to country struggles to attain such standards \parencite{ozawa2018}.
Thus, general approaches are needed that meet real-world data collection.

The paper is organized as follows.
Section \ref{sec:lit} presents related literature regarding PMS and network inference.
Section \ref{sec:dataDescription} describes supply-chain PMS and associated sources of uncertainty.
Section \ref{sec:inferenceProblems} demonstrates the unidentifiability inherent in using PMS testing results.
Section \ref{sec:SFPinferenceResolutions} introduces a Bayesian method for inferring SFP sources.
Section \ref{sec:casestudy} illustrates an application of our method to PMS data from a low- and middle-income country and demonstrates improvements on current PMS practice.
Section \ref{sec:discussion} discusses implementation considerations and future directions.

\vspace{-5pt}
\section{PMS and network inference literature} \label{sec:lit}
\vspace{-5pt}
This section reviews the state of the literature for PMS and network inference in addressing the problem of identifying SFPs in supply chains.

\vspace{-5pt}
\subsection{PMS and SFP detection} \label{subsec:PMSandSFPlit}
Two World Health Organization (WHO) reports from 2017 detailed the global impact of SFPs and highlighted gaps in current monitoring and means of strengthening\reditfin{ the} \beditfin{SFP} regulation\reditfin{ of SFPs}, including PMS \citep{who2017,who2017-2}.
Regulators in\reditfin{ low-resource settings} \beditfin{low- and middle-income countries} face a multitude of challenges: limited operational budgets, overstretched regulatory frameworks, and a global supply chain with little international regulatory coordination.
Procurement streams for many countries involve a web of manufacturers and intermediary suppliers with numerous exchanges before reaching consumers \parencite{unicri2012,usp2020}.
Limited PMS data combined with many potential SFP causes means regulators require more sophisticated analysis tools to better identify SFP sources.

Studies of SFP prevalence span several countries and a variety of pharmaceuticals.
\citet{koczwara2017} analyzed 41 such SFP studies and noted significant differences in SFP prevalence based on sample source, country, and therapeutic class.
\citet{ozawa2018} also described considerable study heterogeneity in a survey of 265 SFP studies.

Current PMS methodologies rely on principles of risk-based surveillance and/or lot-quality assurance sampling.
Risk-based surveillance involves applying regulatory resources as a function of public-health risk and SFP risk.
\citet{nkansah2018} proposed a risk-based PMS approach that maximizes resource utilization in low- and middle-income countries.
\citeauthor{nkansah2018} leveraged resource availability, assessments of SFP risk, and valuations of public-health importance in generating PMS policies.
Risk-based surveillance thus provides guidance on which pharmaceuticals and outlets to sample; lot-quality assurance sampling is a method that provides guidance on the sample sizes required to draw conclusions from PMS data.
\citet{newton2009} developed guidelines for PMS sampling using the lot-quality-assurance-sampling principle of tolerance thresholds for the proportion of pharmaceuticals or outlets of unsuitable quality in a particular region or country.
The regulator sets an SFP tolerance level for each region, and analysis of tested random samples from different regions reveals if the SFP prevalence level within a region exceeds this tolerance.
Risk-based PMS and lot-quality assurance sampling recognize the medicine-specific and regional drivers of SFPs, but upstream supply-chain effects or assessments are not yet fully integrated.

Studies have identified supply-chain factors that drive the generation and distribution of SFPs.
Analysis of falsified products collected throughout sub-Saharan Africa in \citet{newton2011} suggested original manufacture in eastern Asia.
\citet{suleman2014quality} analyzed the impact of supply-chain echelon and other factors in Ethiopia and concluded that the country of manufacturer is the most important indicator for SFPs.
\citet{pisani2019} illustrated how different risks within a pharmaceutical market interact to drive government, industry, counterfeiter and consumer actions using qualitative data from China, Indonesia, Turkey and Romania.
Analyses in \citeauthor{pisani2019} include depictions of how SFPs can be driven by supply-chain factors both inside and outside a given country, with low- and middle-income countries facing more challenges regarding these factors than high-income countries.
The risk-based PMS guidelines of \citet{nkansah2018} acknowledge the effect on SFP prevalence by upstream supply-chain locations in risk calculations but do not use this in analysis of PMS testing data.
 
With recent developments in technology for medical products regulation, there are opportunities for new approaches for PMS sampling and data analysis.
\citet{hamilton2016public} reviewed policies for combating
SFPs under testing uncertainty and called for methodology that accounts for testing accuracy.
The growth of track-and-trace technology, where bar-coded products are followed from manufacturer to outlet, can provide important supply-chain data to improve regulation \parencite{rotunno2014,pisani2019}.
However, the implementation of full track-and-trace systems is resource-intensive.
Low-cost screening tools that supplement expensive and centrally located laboratory testing are well-suited to many\reditfin{ low-resource} \beditfin{low- and middle-income} settings despite their decreased accuracy.
\citet{chen2021} demonstrated that low-cost screening tools have the potential to locate SFPs more cost-effectively than the exclusive use of high-performance laboratory testing.

In summary, supply-chain effects on the occurrence of SFPs are known to be crucial, but these effects are not yet integrated into PMS methodology.
\citet{nkansah2018} used assessments of SFP risk to better allocate limited PMS resources to select consumer-facing sampling locations; we leverage available supply-chain information to extract more analytical power from limited PMS resources.
\citet{newton2009} provided the sampling levels necessary to determine
if SFPs at tested sites exceeded designated threshold rates; the method of this paper provides inference on the SFP rates at tested locations as well as locations upstream in the supply chain.

\subsection{Network inference} \label{subsec:networkInferenceLit}
\vspace{-5pt}
Studies of illicit supply chains span a variety of modeling and solution approaches.
\citet{anzoom2021} reviewed approaches to understanding and disrupting illicit systems.
\citeauthor{anzoom2021} classified studies as taking either a supply-chain view or a network view\reditfin{ to studying these systems}: a supply-chain view models production and distribution processes directly, while a network view considers general associations among\reditfin{ participants} \beditfin{actors}.
For instance, \citet{basu2014} described three supply-chain phases of procurement, concealed transportation, and distribution in the case of wildlife smuggling, while \citet{schwartz2009} proposed measuring nodes in criminal networks according to the nodes' resources and relationships with other nodes.
Our study considering two echelons of a pharmaceutical supply chain falls within the supply-chain view category; \citeauthor{anzoom2021} noted that studies in this category usually meet the context of a particular field rather than generalize to all illicit systems.
Bayesian approaches have also been used for illicit network problems;
\citeauthor{anzoom2021} noted \citet{hussain2008}, which identified principal nodes in criminal social networks, and \citet{triepels2018}, which used shipping documents to detect smuggling.

Our objective is to guide detection of SFP origins given testing at downstream nodes.
This setting belongs to the family of network-inference problems where parameters are determined using \beditfin{measurements from network-deployed sensors at nodes or links}\reditfin{ sensors deployed across a network}.
\reditfin{Network inference, in general, estimates network parameters through measurements taken at nodes or links in a network.}
Nodes can create or store information or products, and a link between two nodes is a possible avenue of traversal of information or products \parencite{diestel2005}.

Network-tomography methods infer unknown network parameters through measurements taken at a subset of network locations \citep{castro2004network}.
Network tomography emerged with the\reditfin{ development of the} internet\beditfin{'s rise}, as\reditfin{ measurements of} transfer delay could only be\reditfin{ taken} \beditfin{measured} at origins and destinations while delay at interior network links remained unknown.
A frequently studied model is $\consolSFPfunc=\transMat\importerSFPrates$, where $\consolSFPfunc$ is a vector of link-level measurements of a phenomenon such as traffic flow or delay, $\importerSFPrates$ is a vector of parameters characterizing phenomena for paths between pairs of nodes, and $\transMat$ is an incidence matrix tying links with paths.
In such models, either $\transMat$ or $\importerSFPrates$ is unknown.
Tomography approaches infer the unknown parameters from data.
The conditions under which network parameters are identifiable under sufficient data are often of interest, so that approaches can be developed that allow\reditfin{ the} \beditfin{parameter} identification\reditfin{ of desired parameters}.
For example, \citet{tebaldi1998bayesian}
considered the problem of inferring road traffic between nodes using link measurements and employed a Bayesian approach to rectify identifiability issues.
Network tomography infers the path-level parameters in $\importerSFPrates$, for example in \citet{chen2010}, or the presence of links in $\transMat$, for example in \citet{ni2010}.

Inference on quality rates in pharmaceutical supply chains parallels prior work in network inference. 
However, to our knowledge, the specific supply-chain structure of untested nodes in a higher echelon that supply tested nodes in a lower echelon cannot be recovered from the structures present in the literature.
A key difference in network inference under PMS is that measurements are expensive, as emphasized by the value of the risk-based approach in \citet{nkansah2018}.
PMS requires obtaining physical samples from pharmaceutical vendors, while network tomography approaches, for instance, can take network measurements every few minutes or seconds \citep{cao2000}.
The strategies to discern parameters in network-inference applications leverage techniques such as Bayesian analysis, distributional assumptions, or
problem-specific characteristics such as user behavior or propagation processes.

\vspace{-5pt}
\section{Modeling supply-chain
PMS} \label{sec:dataDescription}
\vspace{-5pt}
This section describes PMS data collection and types of associated uncertainty.

\vspace{-5pt}
\subsection{Pharmaceutical supply chains} \label{subsec:SCstructure}

The PMS activities\reditfin{ considered in this paper} \beditfin{we study} entail the testing of products sampled from outlets, which are locations where customers purchase products \parencite{nkansah2018}.
We consider the echelon of outlets, plus one upstream echelon shared by outlets.%
\reditfin{An upstream echelon can be a set of manufacturers, aggregates of manufacturers, intermediate distributors who source from manufacturers and disseminate to outlets, or any other supply-chain locations that relate to outlets in the same way.}\reditfin{
Focusing on two connected  echelons within a pharmaceutical supply chain,}
\reditfin{ t}\beditfin{T}he echelon of \emph{test nodes}, denoted by $\outletSet=\{1,\ldots,\numOutlets\}$, is the set of nodes from which the regulator collects samples for testing.
A test node may be an individual seller of pharmaceuticals, or an aggregation of such sellers; \citet{newton2009} considered such aggregates for analysis.
Some echelon from which test nodes source their products is referred to as the echelon of \emph{supply nodes}, denoted by $\importerSet=\{1,\ldots,\numImporters\}$.
Designation of the upstream echelon is a modeling choice left to the regulator and often determined by the metadata available.
For instance, supply nodes may be national importers who procure from international sources\reditfin{ and provide products to local outlets}, or collections of international manufacturers grouped by country of operation. 
This flexibility generalizes to many\reditfin{ low-resource} \beditfin{low- and middle-income} settings, as discussed in Section \ref{sec:discussion}.

Under these definitions, each product passes through exactly one supply node and one test node before collection by a regulator for testing, but products often have passed through other echelons before and after the supply node prior to reaching the test node.
SFP generation at a node may stem from factors merely associated with that node and not because of intrinsic conditions at that node; for instance, an outlet may consistently source from an intermediary injecting falsified products, or a manufacturer may often use a transport service with poor adherence to proper storage conditions.
\reditfin{The approach proposed in this paper}\beditfin{This paper's approach} provides inference on where in the supply chain to further investigate.

\vspace{-10pt}
\subsection{PMS data collection} \label{subsec:PMSsampling}
\vspace{-5pt}
For sample $i$, regulators collect the product from a test node for testing with a binary response:
$\testResultpoint_i=1$ represents SFP detection and $\testResultpoint_i=0$ represents no SFP detection.
We assume collected products are taken uniformly from across all products at the test node, i.e., there is no bias in the SFP probability of the collected product.
This assumption is reasonable as collection occurs before testing, and regulators usually attempt to collect products covertly.
Multiple samples can be collected from each test node.
The test node $\outletLabelpoint_i$ in $\outletSet$ associated with sample $i$ is known at the point of collection, as the regulator visits the test node to collect the sample.
There are two cases for available supply-chain information regarding supply nodes:
\begin{itemize}
    \item \emph{Tracked}: The supply-chain path for each sample is known, meaning sample $i$ includes the supply-node label $\importerLabelpoint_i$ of $\importerSet$.
    For example, the tracked case applies if the supply node is identified on packaging or invoices for samples.
    \item \emph{Untracked}: Instead of knowing the specific supply node-test node path for each sample, the vector of sourcing probabilities from all supply nodes, $\transMat_{\outletLabelpoint}$, is known for each test node $\outletLabelpoint$.
    For example, the untracked case applies if the packaging of samples does not have a supply-node label, but the regulator has access to historical procurement records for test nodes.
    Untracked supply-chain information constitutes the minimum degree of information required to integrate testing data and supply-chain information towards forming inferences.
\end{itemize}
Some supply chains may feature both tracked and untracked elements; however, we generally consider supply chains that are wholly tracked or untracked, and discuss supply chains featuring both information types in Section \ref{sec:discussion}.

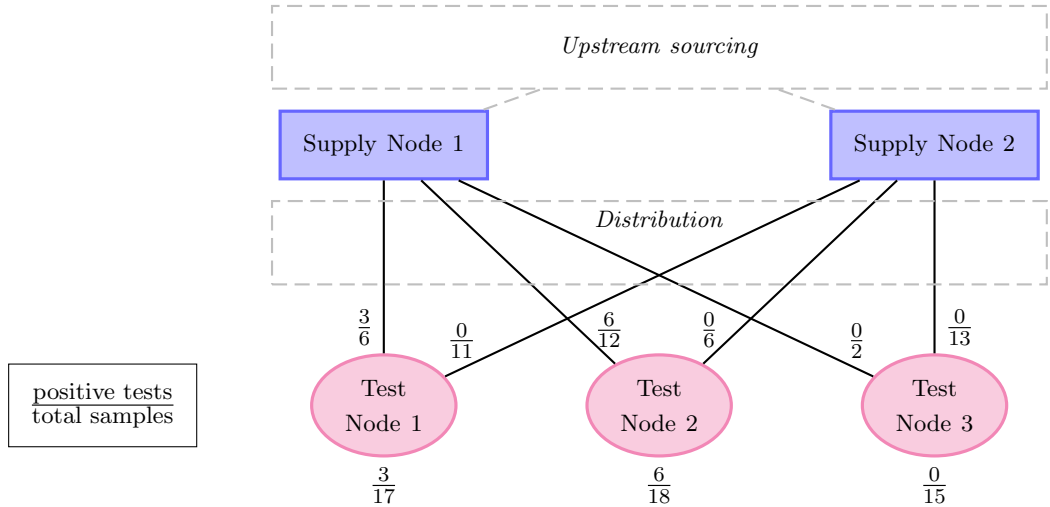
\begin{figure}[t]
    \centering
    \usetikzlibrary{shapes.geometric}
    \begin{tikzpicture}[
    squarednode/.style={rectangle, draw=blue!60, fill=blue!25, very thick,minimum height = 0.9cm,text width=2.5cm, align=center},
    roundnode/.style={ellipse, draw=magenta!60, fill=magenta!25, very thick,  minimum height = 1.3cm, text width=1.1cm, align=center},
    node distance=10mm
    ]
    \node[squarednode](i1){\footnotesize Supply Node 1};
    \node[roundnode, label=below:{\normalsize $\frac{3}{17}$}](o1)[below= 2.3cm of i1] {\footnotesize Test Node 1};
    \node[roundnode, label=below:{\normalsize $\frac{6}{18}$}](o2)[right=1.7cm of o1] {\footnotesize Test Node 2};
    \node[roundnode, label=below:{\normalsize$\frac{0}{15}$}](o3)[right=1.7cm of o2] {\footnotesize Test Node 3};
    \node[squarednode](i2)[above = 2.3cm of o3]  {\footnotesize Supply Node 2};
    \draw[thick,-] (i1) to  node[pos=0.85, left] {\normalsize{$\frac{3}{6}$}} (o1) ;
    \draw[thick,-] (i1) to  node[pos=0.97, above] {\normalsize{$\frac{6}{12}$}} (o2) ;
    \draw[thick,-] (i1) to  node[pos=0.96, above] {\normalsize{$\frac{0}{2}$}} (o3) ;
    \draw[thick,-] (i2) to node[pos=0.96, above] {\normalsize{$\frac{0}{11}$}} (o1);
    \draw[thick,-] (i2) to node[pos=0.97, above] {\normalsize{$\frac{0}{6}$}} (o2);
    \draw[thick,-] (i2) to node[pos=0.85, right] {\normalsize{$\frac{0}{13}$}} (o3);
    \node[draw,left = 1.5cm of o1,minimum height=1.1cm,minimum width=2.5cm] (lgd) {$\frac{\text{\footnotesize positive tests}}{\text{\footnotesize total samples}}$};
    \node[rectangle, thick, draw=lightgray, minimum width = 103mm,text width=25mm, align=center,  minimum height = 0.9cm, anchor=west, dash pattern=on 0.2cm off 0.1cm] (r2) at (-1.5,-1.3) {\footnotesize\textit{Distribution}\vspace{18pt}};
    \node[rectangle, thick, draw=lightgray, minimum width = 103mm,text width=45mm, align=center, minimum height = 1.1cm, anchor=west, dash pattern=on 0.2cm off 0.1cm] (r1) at (-1.5,1.3) {\footnotesize\textit{Upstream sourcing}};
    \draw[thick, -, color=lightgray,dash pattern=on 0.2cm off 0.1cm] (r1) to (i1);
    \draw[thick, -, color=lightgray,dash pattern=on 0.2cm off 0.1cm] (r1) to (i2);
    \end{tikzpicture}
\caption[Extending analysis of PMS test results by one additional echelon]{Extending analysis of PMS test results by one additional upstream echelon.
``Upstream sourcing'' and ``Distribution'' signify supply-chain locations for which information is not considered.}
\label{fig:tracked_example}
\end{figure}

\reditfin{To better explain product flows to the test nodes, an}\beditfin{An} illustrative example in Figure \ref{fig:tracked_example} depicts a tracked supply chain with three test nodes and two supply nodes.
\reditfin{We denote a}\beditfin{A} supply node $\importerLabelpoint$-test node $\outletLabelpoint$ path, also called an $(\outletLabelpoint,\importerLabelpoint)$ arc, \reditfin{as}\beditfin{is} the product route from supply node $\importerLabelpoint$ of $\importerSet$ to test node $\outletLabelpoint$ of $\outletSet$. 
Fraction labels indicate the number of positive tests over the total tests.
\reditfin{Upon collecting testing data, a}\beditfin{A} regulator only inspecting\reditfin{ the} aggregate values at the test-node echelon may conclude that Test Nodes 1 and 2 are significant sources of SFP generation, given their positivity rates.
However, products at these nodes only test positive when sourced from Supply Node 1.
Half of the tested products from Supply Node 1 are SFPs, while no SFPs are associated with Supply Node 2. 
If the test nodes were truly generating SFPs, a more even distribution of discovered SFPs across \reditfin{ the respective}\beditfin{supply-node} paths\reditfin{ of Supply Node 1 and Supply Node 2} would be expected.
It instead seems more reasonable that SFPs stem from upstream factors associated with Supply Node 1.
This example illustrates the importance of supply-chain information for determining SFP sources.

\vspace{-5pt}
\subsection{Sources of uncertainty} \label{subsec:sources_uncertainty}
\vspace{-5pt}
We describe three key sources of uncertainty when inferring SFP sources using PMS.
Fundamental unidentifiability refers to the inability to conclude the origin of
an SFP upon its detection.
Testing accuracy refers to the ability of testing tools to correctly detect SFPs.
Untracked sampling refers to the case where the supply node associated with each test is known probabilistically.

\vspace{-5pt}
\subsubsection{Fundamental unidentifiability} \label{subsubsec:fundamental_unident}
\vspace{-5pt}
There is an inherent inability to identify the sources of SFPs when sampling only at test nodes and not at supply nodes:
it cannot be stated with certainty that SFP generation did or did not occur upstream in the supply chain.

\subsubsection{Testing accuracy} \label{subsubsec:testing_accuracy}
\vspace{-5pt}
Testing tools have an inherent sensitivity and specificity.
Sensitivity refers to the probability of a positive test result given that the tested product is indeed an SFP, and specificity refers to the probability of a negative test result given that the tested product is not an SFP.
\citet{kovacs2014} identified 42 SFP testing technologies and noted sensitivity in the range of 78\%--100\% and specificity in the range of 88\%--100\%, although metrics for some technologies were not reported and testing accuracy can depend on the type of pharmaceutical being tested.
The detected amount of SFPs may\reditfin{ either} increase or decrease from the amount that would be detected with perfectly accurate testing tools, depending on the testing accuracy as well as the SFP rates in the supply chain.
\reditfin{In an environment that is already fundamentally unidentifiable, testing uncertainty further hinders the identification of SFP sources.}

\vspace{-10pt}
\subsubsection{Untracked samples} \label{subsubsec:untracked_samples}
\vspace{-5pt}
In the untracked setting, the supply node associated with a sample\reditfin{ collected at a test node} is unknown. 
It is assumed instead that for each test node, a distribution across supply nodes can be constructed through historical procurement data or other means.
Modeling outlets as test nodes and intermediary distributors as supply nodes, for example, testing data can be integrated with outlet\reditfin{ procurement} records of\reditfin{ prior intermediary} \beditfin{previous} distributor transactions to form untracked PMS data.
The likelihood that test node $\outletLabelpoint$ in $\outletSet$ procures from supply node $\importerLabelpoint$ in $\importerSet$ is\reditfin{ referred to as} \beditfin{called} the sourcing probability of test node $\outletLabelpoint$ from supply node $\importerLabelpoint$, and is captured by element $\transMat_{\outletLabelpoint\importerLabelpoint}$ in matrix $\transMat\in[0,1]^{\numOutlets\times\numImporters}$.  
Note\reditfin{ that} $\transMat$ resembles the path-link incidence matrices reviewed in Section \ref{subsec:networkInferenceLit}.
The row vector corresponding to the set of sourcing probabilities for test node $\outletLabelpoint$ is $\transMat_\outletLabelpoint$.
Thus in the untracked case,\reditfin{ instead of knowing the specific supply node-test node path for each sample,} the sourcing probability vector $\transMat_\outletLabelpoint$ is known for each test node $\outletLabelpoint$.

\vspace{-5pt}
\section{Problems in SFP inference} \label{sec:inferenceProblems}
\vspace{-8pt}
This section defines likelihood functions for PMS data and establishes that tracked and untracked supply chains are unidentifiable.

\subsection{Tracked and untracked likelihoods}
\label{subsec:likelihoods}
\vspace{-5pt}
Binary data $\testResultpoint_1,\ldots,\testResultpoint_\numTests$ are obtained for samples from test nodes in $\outletSet$ that are tested with a device with sensitivity $\diagSens\in[0,1]$ and specificity $\diagSpec\in[0,1]$.
Each sample is collected uniformly from all products at the test node, and products at test nodes are sourced from supply nodes according to sourcing-probability matrix $\transMat$.
Conditional on test node random variable  $\outletLabelRV_i=\outletLabelpoint_i$ for sample $i$, $\importerLabelpoint_i$ is a realization of random variable $\importerLabelRV_i$ which is independently sampled according to the probabilities in row $\transMat_{\outletLabelpoint}$.
The overall set of testing data is represented by $\dataSet=(\bm{\testResultpoint},\bm{\outletLabelpoint},\bm{\importerLabelpoint})$ in the tracked case, where $\bm{\testResultpoint}=\{\testResultpoint_1,\ldots,\testResultpoint_\numTests\}$ is the set of testing results, $\bm{\outletLabelpoint}=\{\outletLabelpoint_1,\ldots,\outletLabelpoint_\numTests\}$ is the set of test-node labels, and $\bm{\importerLabelpoint}=\{\importerLabelpoint_1,\ldots,\importerLabelpoint_\numTests\}$ is the set of supply-node labels.
For the untracked case, $\dataSet=(\bm{\testResultpoint},\bm{\outletLabelpoint},\bm{\transMat})$.
Test-node SFP rates are stored in vector $\outletSFPrates=(\outletSFPrates_1,\ldots,\outletSFPrates_{\numOutlets})\in(0,1)^{\numOutlets}$ and supply-node SFP rates are stored in vector $\importerSFPrates=(\importerSFPrates_1,\ldots,\importerSFPrates_{\numImporters})\in(0,1)^{\numImporters}$.
A node's SFP rate denotes a constant proportion of products traversing the node that become SFP.
As nodes signify real-world locations, SFP rates of exactly $0$ or $1$ are not considered: a rate of $0$ implies a node incapable of error, while a rate of $1$ implies a node only distributing SFPs.

In multi-echelon supply chains, SFPs can be generated at any echelon.
In our modeling of two connected echelons, when we say that products become SFP at either the test node or the supply node, we mean that SFP generation occurs at the test node, at an upstream location associated with the supply node, or at the supply node itself.

The consolidated SFP rate of a sample denotes the probability that the sample is an SFP when accounting for SFP rates at\reditfin{ the} test nodes as well as\reditfin{ the} supply nodes.\reditfin{ For a given supply chain, it} \beditfin{It} suffices to consider only the test node-supply node paths where test nodes have a non-zero probability of sourcing from the supply node.
Let $\coupleSet\subseteq\outletSet\times\importerSet$ be the set of $(\outletLabelpoint,\importerLabelpoint)$ arcs where $\transMat_{\outletLabelpoint\importerLabelpoint}>0$
\reditfin{, such that $|\coupleSet|$ signifies the number of paths from which testing data can be collected in the tracked case}.
The consolidated SFP rate of a tracked sample collected from an $(\outletLabelpoint,\importerLabelpoint)$ arc in $\coupleSet$ is
\setlength{\belowdisplayskip}{5pt} 
\setlength{\belowdisplayshortskip}{5pt}
\setlength{\abovedisplayskip}{5pt} 
\setlength{\abovedisplayshortskip}{5pt}
\begin{equation} \label{eq:tr}
    \consolSFPfunc_{\outletLabelpoint\importerLabelpoint}(\SFPrateSet) = \outletSFPrates_{\outletLabelpoint}  + (1- \outletSFPrates_{\outletLabelpoint}) \importerSFPrates_{\importerLabelpoint} \,.
\end{equation}
The first term of (\ref{eq:tr}) corresponds to the test-node SFP rate and the second term corresponds to the supply-node SFP rate, adjusted for the test-node rate.
This adjustment is necessary as an SFP cannot generate at both the test node and the supply node; we assume once a pharmaceutical is substandard or falsified, additional poor supply-chain conditions do not make the pharmaceutical less suited for consumption.
Further, an SFP cannot be recovered into a non-SFP.
The consolidated SFP rate of an untracked sample collected from test node $\outletLabelpoint$ in $\outletSet$ is
\begin{align}\label{eq:untr}
\consolSFPfunc_\outletLabelpoint(\SFPrateSet) = \sum_{\importerLabelpoint \in \importerSet} \transMat_{\outletLabelpoint\importerLabelpoint} \consolSFPfunc_{\outletLabelpoint\importerLabelpoint}(\SFPrateSet)
= \, &\outletSFPrates_{\outletLabelpoint}   + (1- \outletSFPrates_{\outletLabelpoint}) \sum_{\importerLabelpoint \in \importerSet} \transMat_{\outletLabelpoint\importerLabelpoint} \importerSFPrates_{\importerLabelpoint} \,.
\end{align}
(Note $ \sum_{\importerLabelpoint \in \importerSet} \transMat_{\outletLabelpoint\importerLabelpoint} = 1$ for all $\outletLabelpoint$.)
In the untracked case, each supply node-test node path is weighted according to the sourcing probabilities.

The tracked and untracked contexts differ in the supply-chain information available, yet the expressions of SFP probability are similar.
To simplify notation we use \emph{supply-chain trace} $\elem$ of $\elemSet$ to denote the supply-chain information available at sample collection: $\elem$ of $\elemSet$ is an $(\outletLabelpoint,\importerLabelpoint)$ arc in the tracked case and test node $\outletLabelpoint$ in the untracked case, where $\elemSet$ represents $\coupleSet$ or $\outletSet$, respectively.
The summary of the underlying SFP generation accordingly lies with vector $\consolSFPfunc(\SFPrateSet)$ of length $|\elemSet|$, where element $\consolSFPfunc_\elem(\SFPrateSet)$ of $\consolSFPfunc(\SFPrateSet)$ refers to $\consolSFPfunc_{\outletLabelpoint\importerLabelpoint}(\SFPrateSet)$ for some $(\outletLabelpoint,\importerLabelpoint)$ arc in the tracked case and $\consolSFPfunc_\outletLabelpoint(\SFPrateSet)$ for some test node $\outletLabelpoint$ in the untracked case.
Similarly, supply-chain trace $\elem_i$ associated with sample $i$ refers to either arc $(\outletLabelpoint_i,\importerLabelpoint_i)$ in the tracked case or $\outletLabelpoint_i$ in the untracked case, and random variable $K_i$ is $(\outletLabelRV_i,\importerLabelRV_i)$ in the tracked case or $\outletLabelRV_i$ in the untracked case.

Given sensitivity $\diagSens$ and specificity $\diagSpec$, the probability of a positive SFP test is
$\consolInaccFunc_\elem(\SFPrateSet)=\diagSens\consolSFPfunc_\elem(\SFPrateSet)+(1-\diagSpec)(1-\consolSFPfunc_\elem(\SFPrateSet))$ for each $\elem$ of $\elemSet$.
The random variable $\testResultRV_i$ of test $i$ with supply-chain trace $\beditfin{K_i=}\elem_i$ is 1 with probability $\consolInaccFunc_{\elem_i}(\SFPrateSet)$ and 0 otherwise.
The log-likelihood of $(\SFPrateSet)$ under data $\dataSet$ is
\begin{equation} \label{eq:log-likelihood}
    \ell(\SFPrateSet|\dataSet) = \sum_{i=1}^\numTests
     \bigg[\log[\consolInaccFunc_{\elem_i}(\SFPrateSet)] \testResultpoint_{i} + \log[1-\consolInaccFunc_{\elem_i}(\SFPrateSet)] (1-\testResultpoint_{i})\bigg]\,.
\end{equation}
The log-likelihood has a clearer form when summed over supply-chain traces in $\elemSet$.
Let $\mathcal{I}_\elem=\left\{i\in \{1,\ldots,\numTests\}: \elem_i=\elem \right\}$ be the tests corresponding to $\elem$.
The number of results for $\elem$ is $\numTests_\elem=|\mathcal{I}_\elem|$, with mean positive test rate of $\testPosAvg_\elem=\frac{1}{\numTests_\elem}\sum_{i\in\mathcal{I}_\elem}\testResultpoint_i$.
The log-likelihood in (\ref{eq:log-likelihood}) is equivalently expressed using $\numTests_{\elem}$ and $\testPosAvg_\elem$ as
\begin{equation} \label{eq:log-likelihood2}
    \ell(\SFPrateSet|\dataSet) = \sum_{\elem\in\elemSet} 
     \numTests_\elem \bigg[ \log[{\consolInaccFunc}_{\elem}(\SFPrateSet)] \testPosAvg_{\elem} + \log[ 1-{\consolInaccFunc}_{\elem}(\SFPrateSet)] (1 - \testPosAvg_{\elem}) \bigg] \,. 
\end{equation}
Thus the $\testPosAvg_{\elem}$ and $\numTests_\elem$ values across all $\elem$ in $\elemSet$ are sufficient statistics for the supply-chain traces\reditfin{ observed in} \beditfin{of} the data, as the likelihood can be expressed using these values without other \beditfin{data elements}\reditfin{ dependencies on the data}.
\reditfin{Consequently}\beditfin{As a result}, the likelihood can be computed using a summary of PMS testing results.
A usable PMS summary requires the number of positives and negatives associated with each supply-chain trace.
\reditfin{That this summary is sufficient for establishing the likelihood makes sense when considering that SFP rates are assumed time-invariant; as a result, the order of test results is not important.}

\vspace{-5pt}
\subsection{Unidentifiability} \label{subsec:unidentifiability}
\vspace{-5pt}
The tracked and untracked likelihoods are unidentifiable, i.e., for any set of SFP rates $(\SFPrateSet)$ there exists another set of SFP rates $(\SFPrateSetPrime)$ such that $\ell(\SFPrateSetPrime|\dataSet)= \ell(\SFPrateSet|\dataSet)$.
Unidentifiability means data collection cannot uniquely reveal SFP rates.
Theorems \ref{thrm:tracked} and \ref{thrm:untracked}
state that unidentifiability is assured in the tracked and untracked cases for any set of testing data.
Proofs are in Appendix A.
\begin{thrm}[Tracked unidentifiability]
\label{thrm:tracked}
Let $(\SFPrateSet)$ be any set of SFP rates and let $\dataSet=(\bm{\testResultpoint},\bm{\outletLabelpoint},\bm{\importerLabelpoint})$ be a set of tracked data.
There exists $(\SFPrateSetPrime)\neq(\SFPrateSet)$ such that
\[\ell(\SFPrateSetPrime|\dataSet) = \ell(\SFPrateSet|\dataSet).\]
\end{thrm}

\proofatend
    For original SFP rates $(\SFPrateSet)$ in the tracked setting, we form $(\SFPrateSetPrime)\neq(\SFPrateSet)$ with an initial adjustment of the SFP rate at one test node by some $\epsilon$.
    We use the original rate at this test node and $\epsilon$ to produce adjusted rates at all other nodes that result in $(\SFPrateSetPrime)$ with the same likelihood as $(\SFPrateSet)$.
    
    Let $(\SFPrateSet)$ be any set of SFP rates.
    Select test node $\outletLabelpoint'$ and $\epsilon>0$.
    For each test node $\outletLabelpoint$, set $\outletSFPrates'_{\outletLabelpoint}$ as
    \begin{equation} \label{eq:perturb_TN_tr}
        \outletSFPrates'_{\outletLabelpoint} = \outletSFPrates_{\outletLabelpoint} - \epsilon \frac{1-\outletSFPrates_{\outletLabelpoint}}{1-\outletSFPrates_{\outletLabelpoint'}} \,,
    \end{equation}
    and for each supply node $\importerLabelpoint$, set $\importerSFPrates'_\importerLabelpoint$ as
    \begin{equation} \label{eq:perturb_SN_tr}
        \importerSFPrates'_\importerLabelpoint = \frac{\importerSFPrates_\importerLabelpoint (1-\outletSFPrates_{\outletLabelpoint'}) + \epsilon}{1 - \outletSFPrates_{\outletLabelpoint'} + \epsilon}
        \,.
    \end{equation}
    Inspection of (\ref{eq:perturb_TN_tr}) reveals that a sufficiently small $\epsilon$ assures $\outletSFPrates'_{\outletLabelpoint}>0$ for each $\outletLabelpoint$. 
    For any $\epsilon>0$, inspection of (\ref{eq:perturb_SN_tr}) shows that $\importerSFPrates'_{\importerLabelpoint}<1$, as SFP rates are assumed to be between 0 and 1.
    Thus a sufficiently small $\epsilon$ assures valid adjusted rates $(\SFPrateSetPrime)$ such that $(\SFPrateSetPrime)\neq(\SFPrateSet)$.
    
    Consider the tracked consolidated SFP rate under $(\SFPrateSetPrime)$ for any $(\outletLabelpoint,\importerLabelpoint)$ arc:
    \begin{align*}
         \consolSFPfunc_{\outletLabelpoint\importerLabelpoint}(\SFPrateSetPrime)
         &= \outletSFPrates'_\outletLabelpoint + (1-\outletSFPrates'_\outletLabelpoint) \importerSFPrates'_\importerLabelpoint \\
         &= \outletSFPrates_{\outletLabelpoint} - \epsilon \frac{1-\outletSFPrates_{\outletLabelpoint}}{1-\outletSFPrates_{\outletLabelpoint'}} + \left(1 - \outletSFPrates_{\outletLabelpoint} + \epsilon \frac{1-\outletSFPrates_{\outletLabelpoint}}{1-\outletSFPrates_{\outletLabelpoint'}}\right) \frac{\importerSFPrates_\importerLabelpoint (1-\outletSFPrates_{\outletLabelpoint'}) + \epsilon}{1 - \outletSFPrates_{\outletLabelpoint'} + \epsilon} \\
         &= \outletSFPrates_{\outletLabelpoint} - \epsilon \frac{1-\outletSFPrates_{\outletLabelpoint}}{1-\outletSFPrates_{\outletLabelpoint'}} + \left(\frac{(1-\outletSFPrates_\outletLabelpoint)(1-\outletSFPrates_{\outletLabelpoint'}+\epsilon)}{1-\outletSFPrates_{\outletLabelpoint'}} \right) \frac{\importerSFPrates_\importerLabelpoint (1-\outletSFPrates_{\outletLabelpoint'}) + \epsilon}{1 - \outletSFPrates_{\outletLabelpoint'} + \epsilon} \\
         &= \outletSFPrates_{\outletLabelpoint} + \frac{ \importerSFPrates_\importerLabelpoint(1-\outletSFPrates_{\outletLabelpoint'}) - \outletSFPrates_\outletLabelpoint \importerSFPrates_\importerLabelpoint(1-\outletSFPrates_{\outletLabelpoint'}) - \epsilon + \epsilon\outletSFPrates_\outletLabelpoint + \epsilon - \epsilon\outletSFPrates_\outletLabelpoint }{1-\outletSFPrates_{\outletLabelpoint'}} \\
         &= \outletSFPrates_{\outletLabelpoint} + (1 - \outletSFPrates_{\outletLabelpoint}) \importerSFPrates_\importerLabelpoint = \consolSFPfunc_{\outletLabelpoint\importerLabelpoint}(\SFPrateSet)\,.
    \end{align*}
    Thus $\consolSFPfunc_{\outletLabelpoint\importerLabelpoint}(\SFPrateSetPrime)=\consolSFPfunc_{\outletLabelpoint\importerLabelpoint}(\SFPrateSet)$ for all arcs and $\ell(\SFPrateSetPrime|\dataSet) = \ell(\SFPrateSet|\dataSet)$.
\endproofatend

\begin{thrm}[Untracked unidentifiability]
\label{thrm:untracked}
Let $(\SFPrateSet)$ be any set of SFP rates and let $\dataSet=(\bm{\testResultpoint},\bm{\outletLabelpoint},\transMat)$ be a set of untracked data. 
There exists $(\SFPrateSetPrime)\neq(\SFPrateSet)$ such that
\[\ell(\SFPrateSetPrime|\dataSet) = \ell(\SFPrateSet|\dataSet).\]
\end{thrm}

\proofatend
For original SFP rates $(\SFPrateSet)$ in a supply chain in the untracked setting, we form $(\SFPrateSetPrime)\neq(\SFPrateSet)$ by adjusting the SFP rate at one supply node by some $\epsilon$.
The SFP rates at all test nodes are then adjusted by an amount proportional to the respective sourcing probability of that supply node to produce $(\SFPrateSetPrime)$ with the same likelihood as $(\SFPrateSet)$.

Let $(\SFPrateSet)$ be any set of SFP rates.
Select a supply node $b$ and choose $\epsilon>0$ such that $\transMat_{\outletLabelpoint}(\importerSFPrates-e_b \epsilon)>0$ for test node $\outletLabelpoint$, where $\transMat_\outletLabelpoint$ is the row
vector of sourcing probabilities in $\transMat$ corresponding to $\outletLabelpoint$ and $e_b$ is a vector of length $\numImporters$ with a one at the $b$th element and a zero at all other elements. 
Such an $\epsilon$ exists as all SFP-rates are non-zero.
Set $\importerSFPrates'=\importerSFPrates-e_b \epsilon$.
For each $\outletLabelpoint$, set $\outletSFPrates_{\outletLabelpoint}'$ as
\begin{equation} \label{eq:thrmuntr_perturb}
\outletSFPrates_{\outletLabelpoint}'=\frac{\outletSFPrates_{\outletLabelpoint}(1-\transMat_{\outletLabelpoint}\importerSFPrates)+\epsilon\transMat_{\outletLabelpoint b}} {1-\transMat_{\outletLabelpoint}\importerSFPrates+\epsilon\transMat_{\outletLabelpoint b}}\,.
\end{equation}
As the elements of each row $\transMat_\outletLabelpoint$ sum to one, it follows that $\transMat_\outletLabelpoint\importerSFPrates<1$.
Additionally, since SFP rates and sourcing probabilities are all between zero and one, it holds that $0<\outletSFPrates_\outletLabelpoint'<1$ for any $\epsilon>0$, and thus $(\SFPrateSetPrime)$ are valid rates.
As $\importerSFPrates_\importerLabelpoint'=\importerSFPrates_\importerLabelpoint-\epsilon$ for some $\epsilon>0$, it follows that $(\SFPrateSetPrime)\neq(\SFPrateSet)$.

Consider the untracked consolidated SFP rate under $(\SFPrateSetPrime)$ at any test node $\outletLabelpoint$:
\begin{align*}
    \consolSFPfunc_\outletLabelpoint(\SFPrateSetPrime) &=\outletSFPrates_\outletLabelpoint'+(1-\outletSFPrates_\outletLabelpoint')\transMat_{\outletLabelpoint}\importerSFPrates' \\
    &= \frac{\outletSFPrates_{\outletLabelpoint}(1-\transMat_{\outletLabelpoint}\importerSFPrates)+\epsilon\transMat_{\outletLabelpoint \importerLabelpoint}} {1-\transMat_{\outletLabelpoint}\importerSFPrates+\epsilon\transMat_{\outletLabelpoint \importerLabelpoint}} + \frac{(1-\outletSFPrates_{\outletLabelpoint})(1-\transMat_{\outletLabelpoint}\importerSFPrates)} {1-\transMat_{\outletLabelpoint}\importerSFPrates+\epsilon\transMat_{\outletLabelpoint \importerLabelpoint}} \transMat_\outletLabelpoint (\importerSFPrates - e_\importerLabelpoint \epsilon) \\
    &= \frac{\outletSFPrates_{\outletLabelpoint}(1-\transMat_{\outletLabelpoint}\importerSFPrates) + \transMat_\outletLabelpoint\importerSFPrates - \transMat_\outletLabelpoint\importerSFPrates(\transMat_\outletLabelpoint\importerSFPrates - \epsilon\transMat_{\outletLabelpoint\importerLabelpoint}) - \outletSFPrates_\outletLabelpoint(\transMat_\outletLabelpoint\importerSFPrates-\epsilon\transMat_{\outletLabelpoint\importerLabelpoint}) + \outletSFPrates_\outletLabelpoint\transMat_{\outletLabelpoint}\importerSFPrates (\transMat_\outletLabelpoint\importerSFPrates-\epsilon\transMat_{\outletLabelpoint\importerLabelpoint}) } {1-\transMat_{\outletLabelpoint}\importerSFPrates+\epsilon\transMat_{\outletLabelpoint \importerLabelpoint}} \\
    &= \frac{\outletSFPrates_{\outletLabelpoint} + (1-\outletSFPrates_\outletLabelpoint)\transMat_{\outletLabelpoint}\importerSFPrates - \transMat_\outletLabelpoint\importerSFPrates[\outletSFPrates_{\outletLabelpoint} + (1-\outletSFPrates_\outletLabelpoint)\transMat_{\outletLabelpoint}\importerSFPrates] + \epsilon\transMat_{\outletLabelpoint\importerLabelpoint}[\outletSFPrates_{\outletLabelpoint} + (1-\outletSFPrates_\outletLabelpoint)\transMat_{\outletLabelpoint}\importerSFPrates]} {1-\transMat_{\outletLabelpoint}\importerSFPrates+\epsilon\transMat_{\outletLabelpoint \importerLabelpoint}} \\
    &= \outletSFPrates_{\outletLabelpoint} + (1-\outletSFPrates_\outletLabelpoint)\transMat_{\outletLabelpoint}\importerSFPrates = \consolSFPfunc_\outletLabelpoint(\SFPrateSet)\,.
\end{align*}
Thus $\consolSFPfunc_\outletLabelpoint(\SFPrateSetPrime)=\consolSFPfunc_\outletLabelpoint(\SFPrateSet)$ for all test nodes and $\ell(\SFPrateSetPrime|\dataSet) = \ell(\SFPrateSet|\dataSet)$.
\endproofatend

\noindent
Establishing unidentifiability in supply-chain PMS is a core contribution.
We show unidentifiability exists when only considering two echelons of a supply chain; a corollary is that consideration of additional echelons \beditfin{also} implies unidentifiability challenges\reditfin{, as well}.
Thus, SFP rates\reditfin{ throughout a supply chain} cannot be recovered through PMS as currently practiced.
Unidentifiability indicates a need for approaches that distinguish among multiple explanations for a set of data; Section \ref{sec:SFPinferenceResolutions} presents such an approach.

\vspace{-7pt}
\section{SFP-inference resolution}
\label{sec:SFPinferenceResolutions}
\vspace{-6pt}
Theorems \ref{thrm:tracked} and \ref{thrm:untracked} show that identification of unique SFP rates explaining PMS data is not possible; yet, unidentifiability does not eliminate prospects for inferring SFP sources.
This section presents a Bayesian approach to statistical inference of SFP rates that mitigates identifiability issues.

\vspace{-5pt}
\subsection{Bayesian mitigation of unidentifiability} \label{subsec:postDensity}
\vspace{-5pt}
Bayesian analysis combines observations and prior beliefs to infer unknowns.
Placing priors on $(\SFPrateSet)$ encodes beliefs about SFP generation that distinguish candidate SFP rates with similar likelihoods.
For example, \citet{tebaldi1998bayesian}
employed a Bayesian approach to alleviate identifiability issues for pair-wise traffic counts for nodes in a network.
Given different vectors of SFP rates with similar likelihoods under a set of PMS data and supply-chain information, prior expectations of the level and dispersal of SFPs across the supply chain help discern plausible vectors of SFP rates.

Let $\priorFunc(\SFPrateSet)$ be a prior density on $(\SFPrateSet)$.
Multiplying $\priorFunc(\SFPrateSet)$ with the likelihood under data $\dataSet$, $\exp{(\ell(\SFPrateSet | \dataSet))}$, is then proportional to the posterior, i.e.,
\begin{equation} \label{eq:posterior}
 \postFunc(\SFPrateSet | \dataSet) \propto \exp (\ell(\SFPrateSet | \dataSet)) \priorFunc(\SFPrateSet)\,.
\end{equation}
\reditfin{If the posterior is concentrated}\beditfin{Posterior concentration} at a region of high SFP rates for a particular node\reditfin{, testing data and supply-chain} \beditfin{means that available}  information indicate that node\reditfin{ is} \beditfin{as} a credible SFP source.
\reditfin{If the posterior is concentrated}\beditfin{Posterior concentration} at a region of low SFP rates\reditfin{,} \beditfin{indicates} that node is not a credible SFP source.
\reditfin{If the posterior is not concentrated anywhere, then}\beditfin{Non-concentration of the posterior means} data are insufficient to overcome sources of uncertainty.%
\reditfin{Thus the posterior in (6) can reveal nodes that have credible associations with high or low SFP generation rates as well as those nodes that require further testing.}

Sections \ref{subsec:priorFormation} and \ref{subsec:usingMCMC} discuss prior formation and generating suitable \beditfin{posterior} draws\reditfin{ from the posterior}.
Sections \ref{subsec:inferenceExample} and \ref{subsec:interpretPostSamps} first illustrate the application of inference in a PMS context.

\vspace{-8pt}
\subsection{Inference example} \label{subsec:inferenceExample}
\vspace{-5pt}
We revisit the example from Section \ref{subsec:PMSsampling} from a Bayesian perspective.
Suppose one\reditfin{ has a prior belief} \beditfin{believes} that SFP rates at nodes are independent and, while\reditfin{ they} \beditfin{nodes} could exhibit SFP rates near 40\%, most nodes will exhibit SFP rates below 20\%.
A prior that meets this criteria on test-node SFP rates $\outletSFPrates=(\outletSFPrates_1,\outletSFPrates_2,\outletSFPrates_3)$ and supply-node SFP rates $\importerSFPrates=(\importerSFPrates_1,\importerSFPrates_2)$ is
\begin{equation*}
    \priorFunc(\SFPrateSet) \propto \prod\limits_{\outletLabelpoint\in \{1,2,3\}} \exp\left\{-\frac{1}{2}\left[\logitfunc(\outletSFPrates_\outletLabelpoint)+2\right]^2\right\} \prod\limits_{\importerLabelpoint\in \{1,2\}} {\exp\left\{{-\frac{1}{2}\left[\logitfunc(\importerSFPrates_\importerLabelpoint)+2\right]^2}\right\}},
\end{equation*}
where $\logitfunc(x)=\log(\frac{x}{1-x})$ is the logit function.
Using the logit transformation moves analysis to the real number line\reditfin{;}\beditfin{:} manipulation of the posterior on the real number line avoids computational issues that arise as SFP rates approach zero or one. 

\begin{figure}
    \centering
    \subfloat{
    \includegraphics[width=0.55\textwidth]
    {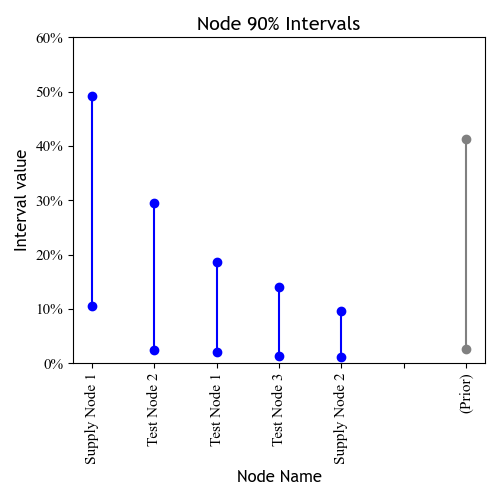}
    }
    \caption{5\% and 95\% quantiles for the posterior of example testing data in Figure \ref{fig:tracked_example}.
    } \label{fig:mcmcexample}
\end{figure}

Combining the prior with the likelihood under the testing data from Figure \ref{fig:tracked_example} yields the posterior.
Figure \ref{fig:mcmcexample} depicts the 5\% and 95\% quantiles for 1,000 posterior draws of $(\SFPrateSet)$.
The quantile\beditfin{s}\reditfin{ information} for the prior\reditfin{ is} \beditfin{are} included for reference.
A node associated with sufficiently high 5\% quantiles indicates a significant posterior probability that SFP generation is linked with that node.
For instance, Supply Node 1 likely constitutes a high SFP risk.
However, although\reditfin{ nine of twenty} \beditfin{9 of 20} tests associated with Supply Node 1 are SFPs (45\%), the prior and the\reditfin{ possibility} \beditfin{chance} that the test nodes are responsible for some SFP generation mean that most weight for the interval for Supply Node 1 falls below the raw percentage of 45\%.
A high 95\% quantile means that \beditfin{sufficient data may show} the associated node\reditfin{ could} \beditfin{to} be a large driver of SFPs\reditfin{; however, insufficient testing data exist to be confident that the node is linked with high SFP levels}.
For example, Test Node 2 has a 95\% quantile near 30\%\reditfin{;}\beditfin{:} more data collection may plausibly show that Test Node 2 is associated with a higher SFP rate than Supply Node 1.

\vspace{-10pt}
\subsection{Interpreting posterior samples} \label{subsec:interpretPostSamps}
\vspace{-5pt}
Draws from the posterior are used to build credible regions for the values of $(\SFPrateSet)$ that generated the data $\dataSet$.
Credible regions signify a space of SFP rates with $1-\alpha$ posterior probability, for some desired $\alpha$ level.
For example, the 5\% and 95\% quantiles are used to build a 90\% interval.
Wide intervals for particular nodes indicate that data are insufficient to draw conclusions.
Drivers of inconclusive intervals include low sample size and the uncertainty sources of Section \ref{subsec:sources_uncertainty}.

Interval interpretation should consider at least three categories for the application of regulatory resources.
Similar to the thresholds of the lot-quality assurance sampling approach of \citet{newton2009}, categorization of nodes along the lines of ``act,'' ``do not act,'' and ``gather more data before deciding,'' allows regulators to translate PMS results into the allocation of intervention resources or further PMS activities.
Categorization aids efficient use of limited resources under uncertainty.

We suggest using acceptance thresholds to build categories from posterior intervals.
The first category includes nodes with interval lower bounds above some lower threshold $l$, where $l$ signifies an SFP rate that triggers the use of further intervention resources.
Data are sufficient to suggest that the SFP rates associated with members of this first category are as high as $l$.
If the example in Figure \ref{fig:mcmcexample} uses 
$l=5\%$, then Supply Node 1 is categorized as a high SFP risk.
Designation of $l$ by regulators should consider the availability of intervention resources as well as what SFP rates are unacceptable in their domains.
For instance, \citet{newton2009} noted WHO guidelines for malaria programs that suggest a change in policy once treatment failure exceeds 10\%; similar treatment-specific rates may guide designation of $l$ for different pharmaceuticals. 

The second category includes nodes with interval lower bounds below $l$ but upper bounds above an upper threshold $u$.
SFP rates for members of this category are potentially as high as $u$, but more data are required to assert that SFP rates are not below $l$.
Thus, targeting further PMS sampling of these nodes may be recommended.
If the example in Figure \ref{fig:mcmcexample} uses $l=5\%$ and $u=20\%$, then Test Node 2 is a moderate SFP risk.
Designation of $u$ by regulators should consider what additional resources can be expended in investigating nodes with the potential for high SFP rates: setting $u$ too low means potentially categorizing all nodes as moderate SFP risks.

The third category captures nodes associated with intervals that have upper bounds below $u$ and lower bounds below $l$.
Nodes in this category are\reditfin{ less likely than nodes in the first category} \beditfin{least likely} to pose significant SFP risk.

\vspace{-8pt}
\subsection{Prior formation} \label{subsec:priorFormation}
\vspace{-3pt}
A variety of prior forms can be used with (\ref{eq:posterior}).
Effective priors encode regulator expectations of SFP generation with respect to size, variability, and dispersal pattern.
Priors are beneficial \beditfin{for mitigating unidentifiability} when informed by reliable \beditfin{regulatory} domain knowledge\reditfin{: regulatory experience is crucial to conducting inference under unidentifiability}.

Applications including the modeling of movie sales and\reditfin{ the studying of} interventions against\reditfin{ sexually transmitted} infections have employed density transformations to enable application-specific analysis \parencite{ainslie2005,hui2020}.
Similarly, an effective strategy here is developing priors on the real number line and transforming the resulting posterior SFP rates to the $(0,1)$ interval for analysis.
Priors defined on the real number line also favorably correspond with the SFP rates indicated by studies in the literature: the resulting distributions have long tails, which aligns with the heterogeneity of SFP generation noted by \citet{who2017-2}.

Consider an independent normal prior, expressed as
\[
\priorFunc\big(\SFPrateSet\big) \propto \prod\limits_{\outletLabelpoint\in \outletSet} \exp\left\{-\frac{1}{2}\left[\frac{\logitfunc(\outletSFPrates_\outletLabelpoint)-\expitConstantMean}{\expitConstantVar}\right]^2\right\} \prod\limits_{\importerLabelpoint\in \importerSet} {\exp\left\{{-\frac{1}{2}\left[\frac{\logitfunc(\importerSFPrates_\importerLabelpoint)-\expitConstantMean}{\expitConstantVar}\right]^2}\right\}}\,.
\]
Parameter $\expitConstantMean$ signifies a prior belief of the standard SFP rate at test nodes and supply nodes, and $\expitConstantVar$ corresponds to SFP-rate spread.
The standard parameter centers expectations of SFP prevalence throughout the supply chain.
The spread parameter reflects the anticipated variety across rates.
For example, a normal prior on the real number line with $\expitConstantMean=-2$ and $\expitConstantVar=1$ produces a distribution in the $(0,1)$ space with respective 5\%, 50\% and 95\% quantiles of 3\%, 12\% and 41\%.

For low $\expitConstantMean$ values, the independence within each prior reflects an assumption that it is unlikely that many SFP sources exist: one node carrying an SFP rate above $\expitConstantMean$ has higher prior likelihood than many nodes carrying such SFP rates.
Using priors with lower spread parameters requires more testing data to pull the posterior probability towards regions favored by the likelihood.

Consider an independent Laplace prior, which carries a similar shape to the normal:
\[
\priorFunc\big(\SFPrateSet\big) \propto \prod\limits_{\outletLabelpoint\in \outletSet}{\exp\left\{{-\frac{|\logitfunc(\outletSFPrates_\outletLabelpoint)-\expitConstantMean|}{\expitConstantVar}}\right\}} \prod\limits_{\importerLabelpoint\in \importerSet}{\exp\left\{{-\frac{|\logitfunc(\importerSFPrates_\importerLabelpoint)-\expitConstantMean|}{\expitConstantVar}}\right\}}\,.
\]
For average and spread similar to the normal, an independent Laplace concentrates nearer the average and has heavier tails.
A Laplace prior reflects\reditfin{ stronger} \beditfin{an} anticipation that some nodes will have SFP rates far from the average; thus the Laplace may better suit consideration of falsification, where falsifiers exploit available yet limited entry points \parencite{who2017-2}.
A normal prior reflects\reditfin{ stronger} \beditfin{an} expectation that rates will vary nearer the average; thus the normal may better suit\reditfin{ consideration of} substandardization\reditfin{, as}\beditfin{:} production, transportation and storage\reditfin{ of pharmaceuticals} entail\reditfin{s} similar activities conducted by different actors.

\vspace{-5pt}
\subsection{MCMC sampling} \label{subsec:usingMCMC}
\vspace{-5pt}
To build the inference described in Section \ref{subsec:interpretPostSamps}, samples from the posterior are needed.  
The posterior in (\ref{eq:posterior}) does not exhibit natural sampling, but\reditfin{ modern} tools such as Markov chain Monte Carlo (MCMC) allow sampling from general posteriors.
Our study uses the No-U-Turn Sampler (NUTS) sampler of \citet{hoffman2014no}.
This sampler uses posterior gradient information;
Supplementary Material I contains applicable posterior derivatives.
The NUTS sampler requires a number of samples to warm start, as well as a parameter, $\delta$, that governs how the algorithm proposes samples.
Our analysis uses $\delta$ of 0.4, which falls within the region suggested by \citeauthor{hoffman2014no}.
The analysis then generates 5,000 warm-start draws and 1,000 draws for inference; more inference draws could be used, but 1,000 draws appear sufficient for\reditfin{ the} analysis
(see Supplemental Material II).

Computation time is not a major restriction for analyzing data common to many low- and middle-income settings.
Supplementary Material II
describes\reditfin{ significant} drivers of computation time.
In general, more nodes increases the dimensionality of $(\SFPrateSet)$ and slows down\reditfin{ the sampler} \beditfin{sampling}.
However,\reditfin{ the overall scale of} computation time for a system with a hundred nodes is\reditfin{ in} seconds, and supply chains in most cases will not feature more than a few hundred nodes.
Our code is publicly available
\if0\blind{on Github as Python package \texttt{logistigate} \parencite{logistigate2021}}\fi
\if1\blind{as [\textit{software removed for blind review}]}\fi.

\vspace{-5pt}
\section{Case study} \label{sec:casestudy}
\vspace{-5pt}
Several national regulatory agencies in low- and middle-income countries provide data to United States Pharmacopeia's Medicines Quality Database (MQDB) to strengthen global regulatory capacity.
We use a PMS data set from MQDB to show how incorporating upstream information can add to the understanding of SFP sources.
The case study demonstrates unidentifiability in real PMS data and shows the value of our Bayesian approach over current practice.

\vspace{-10pt}
\subsection{Case-study setting} \label{subsec:datadescription}
\vspace{-5pt}
The data consist of products collected and tested by a country's pharmaceutical regulatory agency in 2010.
The data are anonymized to protect the country's sources and mask the outlets and manufacturers involved.
A data record denotes purchasing and testing information for a single form of a pharmaceutical product as sold to consumers, e.g., a box of 12 tablets.
A test result is either ``Pass,'' meaning compliance with registration specification, or ``Fail,'' meaning non-compliance.
Each record is associated with multiple geographic divisions.
We consider the ``District'' and ``Manufacturer'' levels of the supply chain, where District refers to the second-largest geographic sub-division of the country.
We model Manufacturers as supply nodes and Districts as test nodes.

The case-study data feature 25\reditfin{ unique} Manufacturers and 23\reditfin{ unique} Districts \beditfin{in 406 PMS records}.\reditfin{ An arc exists between two supply-chain locations if there is a test in the data set bearing the labels of both locations.
For the case study, there are 165 observed District-Manufacturer arcs.}
The\beditfin{se} data contain\reditfin{ 406 PMS records, including} 73 positive tests, or an 18\% SFP rate.
An 18\% SFP rate suggests significant quality issues for the areas sampled by regulators; however, examination of the testing results by only supply-node or test-node label reveals difficulties in defining SFP sources.
District 8 features 7 SFPs of 12 associated tests (58\%), District 7 features 24 SFPs of 81 tests (30\%), and District 16 features 8 SFPs of 44 tests (18\%).
A natural regulatory response would be to dedicate intervention resources to these districts with SFP rates exceeding the national average of 18\%.
At the same time, 8 of 21 samples from Manufacturer 8 are SFPs (38\%), 28 of 92 samples from Manufacturer 5 are SFPs (30\%), and 5 of 31 samples from Manufacturer 3 are SFPs (16\%).
Examining data by manufacturer, a justifiable regulatory response would be to dedicate resources to investigating supply-chain factors associated with these manufacturers.

\vspace{-10pt}
\subsection{Limitations of current methods} \label{subsec:benchmarkComparison}
\vspace{-5pt}
Lot-quality assurance described in \citet{newton2009} uses standard 90\% confidence intervals for proportions to determine if SFP prevalence for a node exceeds quality thresholds, where the 90\% interval for a given proportion $\hat{z}$ and number of samples $n_{\hat{z}}$ is given as $\hat{z} \pm 1.645\sqrt{{\hat{z}(1-\hat{z})}/{n_{\hat{z}}}}$.
The proportion $\hat{z}$ can relate to either a test node or a supply node.
For instance, the standard interval for Manufacturer 13 is $(7\%, 22\%)$ and the standard interval for District 5 is $(5\%, 25\%)$.
The intervals for six districts and eight manufacturers exceed a threshold of $l=5\%$.
A typical regulator response would be to allocate investigative and intervention resources to these locations.

Additionally, a common requirement for the standard interval is $n_{\hat{z}}\hat{z}\geq 5$ and $n_{\hat{z}}(1-\hat{z})\geq 5$ \parencite{mann2010}.
In this data set, the requirement is satisfied by only 5 of 25 manufacturers and 6 of 23 districts.
Obtaining sufficient tests for all test and supply nodes may be infeasible in many resource-limited settings, as in our experiences at \if1\blind [\textit{redacted organization}]%
\fi \if0\blind USP%
\fi: regulators must often allocate resources using insufficient\reditfin{ levels of test results} \beditfin{numbers of tests}.
In contrast, our method does not have a minimum to complete inference; Manufacturer 2, for example, is featured on only one test.

\vspace{-10pt}
\subsection{Manufacturer-District analysis} \label{subsec:casestudyanalysis}
\vspace{-5pt}
The expectations grounding the prior in the Manufacturer-District analysis follows past work in PMS.
Previous studies, such as those reviewed in \citet{ozawa2018}, typically reported aggregated rates across countries, geographic regions, or sub-divisions of the pharmaceutical market.
Research shows that although SFPs are a widespread global problem, SFP generation is heterogeneous: much of the supply chain exhibits low rates while many SFPs derive from a few supply-chain locations \parencite{who2017-2, unicri2012}.
Thus the prior for analysis employs an average SFP rate anchored to what previous studies indicate and a spread sufficiently high to capture anticipated heterogeneity.
An independent Laplace prior with average $\expitConstantMean=-2.5$ and spread parameter $\expitConstantVar=1.3$ produces an average of 15\%, a median of 8\%, and a 90\% interval of 
[0.4\%, 62\%], meaning the prior carries a long right tail covering high SFP-rate regions.
70\% of prior weight falls below an SFP rate of 14\%.
The sensitivity analysis in 
Supplementary Material III
indicates that prior choice does not have an instrumental effect on interval width; sufficient data seem to counterbalance the prior designation.

The Manufacturer-District analysis assumes perfect testing sensitivity and specificity to enable easier isolation of the effects of fundamental unidentifiability and untracked settings.
As expected, sensitivity analysis shows that testing-tool uncertainty generally has an inflationary effect on inference; this inflationary effect is larger for nodes for which there are less data.

\begin{figure}
    \centering
    \subfloat{\includegraphics[width=0.95\textwidth]{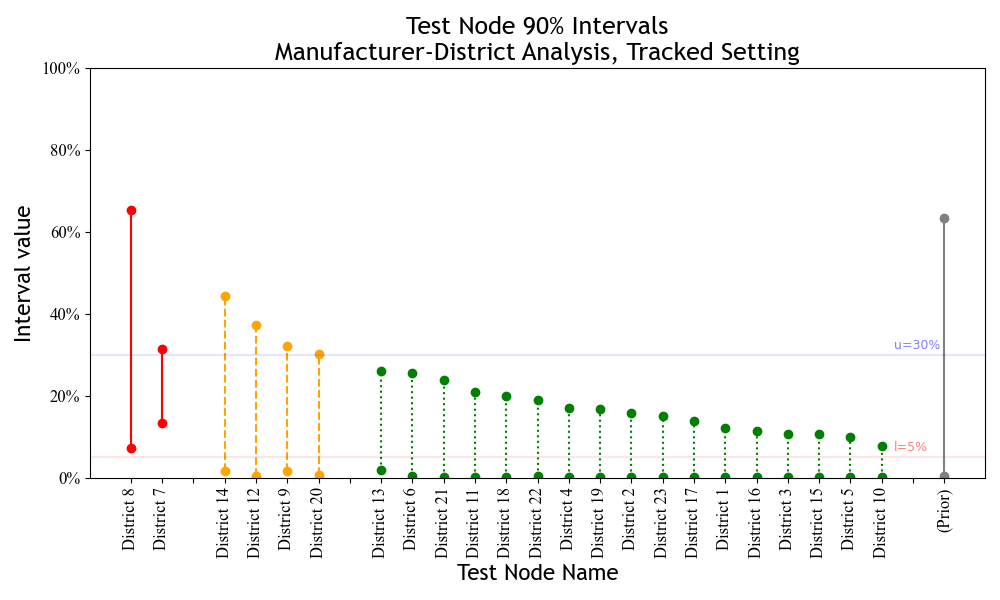} }
    \\
    \subfloat{\includegraphics[width=0.95\textwidth]{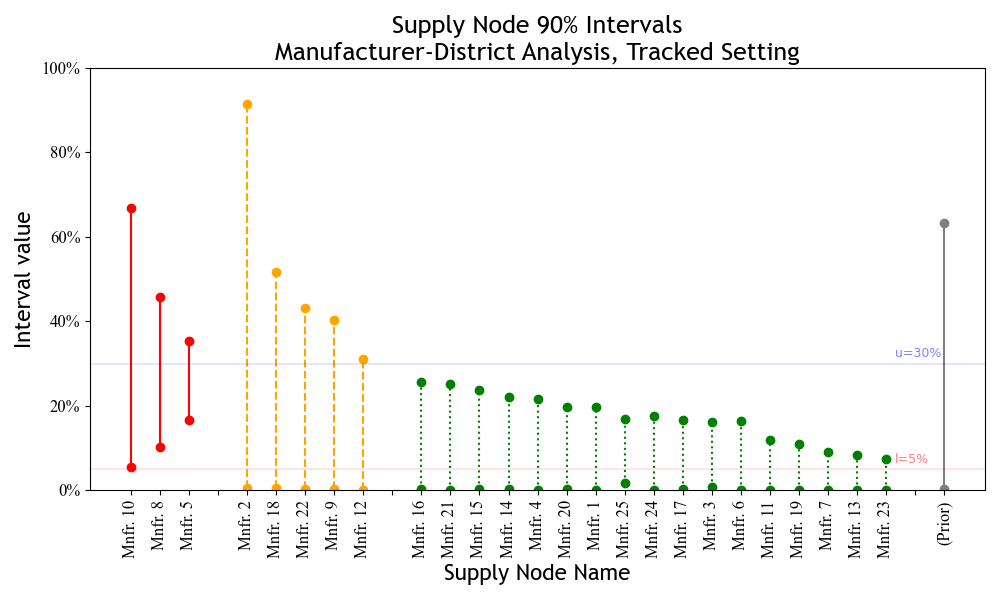}}
    \caption{Test-node \beditfin{and supply-node} 90\% intervals for MCMC samples generated using case-study data \beditfin{in the tracked setting}.
    Intervals with lower bounds above $l=5\%$ are featured in solid lines on the left, intervals with lower bounds below $l$ and upper bounds above $u=30\%$ are featured in dashed lines in the middle, and all other intervals are featured in dotted lines on the right.
    } \label{fig:manudist_TN}
\end{figure}

Figures 3 and 4
show 90\% intervals for SFP rates corresponding to Districts and Manufacturers, respectively, under both tracked and untracked settings.
\beditfin{(Figure 4 is in Appendix B.)}
The figures use the classification scheme described in Section \ref{subsec:interpretPostSamps} with $l=5\%$ and $u=30\%$.
First consider the tracked setting.
Our method's credible intervals are comparable in width with the standard intervals described in Section \ref{subsec:benchmarkComparison}; however, as SFP rates are not considered across the supply chain, the standard intervals are shifted higher by ten to thirty percentage points.
Let the raw SFP rate of a given node be the SFP rate of all tests associated with that node, despite the upstream or downstream supply-chain factors causing SFPs.
Raw rates only apply to supply nodes in the tracked case, as the supply node for each test is unknown in the untracked case.
The raw SFP rates for samples from Manufacturers 10, 8, and 5 are 57\%, 38\%, and 34\%, respectively, and the raw rates associated with District 8 and 7 are respectively 58\% and 30\%.
The raw SFP rates sit near the interval upper bounds for each Manufacturer and District; the interval upper bound for Manufacturer 5, for instance, is 38\%.
The intervals skew lower than the raw rates for all nodes.
The posterior constituting these intervals is accounting for a prior with a low average, in addition to the possibility that SFPs are generated at either test nodes or supply nodes.
\beditfin{In addition, our method reflects uncertainty from low levels of data.
For instance, Manufacturer 2 has only one associated (positive) test: the associated interval spans most of $(0\%,100\%)$.}

Direct consideration of\reditfin{ available} supply-chain connections and\reditfin{ the} associated testing data lends credence to the inferences illustrated in the figures.
All seven\reditfin{ of the} SFPs associated with District 8 are tied to Manufacturers with at least 15\% raw SFP rates: Manufacturers 3, 5, 8, 10 and 24.
The data also feature four non-SFPs for tests from the District 8-Manufacturer 13 arc, which does not support District 8 as a major SFP source.
District 7, on the other hand, is associated with 19 tests featuring Manufacturer 13, where 8 of these tests are SFPs (42\%).
The other 43 tests involving Manufacturer 13 feature only one SFP; thus, District 7 is likely a significant SFP source.
The standard interval for Manufacturer 13 is $(7\%, 22\%)$, while our method's interval for Manufacturer 13 is (0.1\%, 9\%). 
A regulator using the standard interval under $l=5\%$ would find Manufacturer 13 to be associated with significant SFP sources.
Thus, we observe how the posterior addresses the challenge of fundamental unidentifiability by integrating testing data, supply-chain information, and prior expectations to create credible intervals that regulators can use to improve policy decisions.

The analysis also illustrates the importance of supply-chain connections for forming inferences in the tracked setting: without interconnected nodes, fundamental unidentifiability renders too many SFP scenarios as plausible.
Sourcing patterns limit the number of scenarios that can credibly explain the data.
The interval associated with District 5 is an example of the importance of sourcing patterns.
Although testing at District 5 yields 5 SFPs in 34 samples (15\%), District 5 has a narrower interval than the interval for District 17, which yields no SFPs in 9 samples.
Inspection of the manufacturers associated with District 5 samples reveals that all 5 SFPs are sourced from Manufacturer 5.
The standard interval for District 5 is $(5\%, 25\%)$, exceeding the lower threshold of $l=5\%$, while our method's credible interval for District 5 is (2\%, 9\%).
Instead of suspecting SFP generation at District 5, supply-chain information allows us to infer the opposite: District 5 is less likely an SFP source than another test node, District 17, with no detected SFPs.
A regulator using standard intervals may invest intervention resources in District 5, whereas incorporating supply-chain information avoids this investment.
Thus, the inferences resulting from our approach can help regulators determine if data are sufficient to invest limited regulatory resources.

For the\reditfin{ same data under an} untracked setting, the sourcing-probability matrix, $\transMat$, is used as the supply-chain trace instead of the supply-node labels. 
The estimated element of $\transMat$ corresponding to District $\outletLabelpoint$ and Manufacturer $\importerLabelpoint$ is formed by dividing the number of observed samples from arc $(\outletLabelpoint,\importerLabelpoint)$ by the total number of samples collected from test node $\outletLabelpoint$.
The resulting matrix is sparse:\reditfin{ as} test nodes only source from a subset of\reditfin{ all observed} supply nodes.
Comparing the \beditfin{tracked and untracked} intervals\reditfin{ under tracked and untracked settings}, as shown in 
Figures 3 and 4,
reveals the value of tracked over untracked information.
The intervals associated with test nodes remain nearly identical, while the intervals associated with supply nodes change considerably.
This effect is reasonable: we know test nodes exactly and supply nodes only probabilistically.
\reditfin{Under the untracked setting, the }\beditfin{The untracked} supply-node intervals still indicate that \beditfin{upstream supply-chain} factors\reditfin{ upstream in the supply chain} are associated with SFP generation, as shown by the many Manufacturers classified as moderate risks.
\reditfin{Inferring}\beditfin{However, inferring} the most critical upstream direction from many options is unclear.
Untracked analysis for this case study thus carries Type II risk, where potential upstream sources of SFP are missed.

In the untracked case, the structure of $\transMat$ is an important factor in the ability to overcome unidentifiability.
In particular, untracked inference is hampered when test nodes possess similar sourcing patterns.
For instance, 18 of 23 Districts had more than 10\% of associated tests tied to Manufacturer 5.
SFPs associated with upstream supply-chain factors become difficult to infer. 
Even if it is known that upstream factors are principal SFP drivers,
SFPs can just as likely stem from a supply node with high market share and a low SFP rate as from a supply node with low market share and a high SFP rate.
Accordingly, the ideal sourcing environment for successful untracked inference is \beditfin{an environment} where each test node sources from a small subset of supply nodes, with only a few shared supply nodes among any subset of test nodes.

Another challenge to untracked inference is sufficiently estimating the sourcing structure from past data.
Estimating $\transMat$ from procurement or sourcing records carries variance due to the sampling variance of the records.
Supplementary Material III
examines inference sensitivity to the estimation of $\transMat$ using bootstrap sampling; use of different $\transMat$ estimates for this case study impacts the resulting inference for supply nodes but not for test nodes.\reditfin{In sum, use of untracked data carries inherent challenges in detecting upstream SFP sources, particularly if supply-chain information for estimating $\transMat$ is too limited.}
\beditfin{In sum, untracked settings carry challenges for inferring upstream SFP sources, particularly if information for estimating $\transMat$ is too limited.}

\vspace{-5pt}
\section{Conclusion and discussion} \label{sec:discussion}
\vspace{-5pt}
Regulators in low- and middle-income countries can benefit from new tools and methods to maximize the power of surveillance activities.
This paper characterizes the challenge of identifying SFP sources under PMS and demonstrates how the analytical capacity of PMS can be expanded by consideration of supply-chain information.
Our case study illustrates how a Bayesian approach can be combined with domain expertise through well-chosen priors to strengthen identification of SFP sources.
PMS data, including upstream supply-chain information, are already collected routinely; this paper provides a means of extracting more utility from this regular activity.

In addition to limited budgets, the WHO has identified poor international coordination as a significant challenge for regulators in low- and middle-income countries \parencite{who2017-2}.
Placing PMS in a supply-chain context opens avenues for collaboration among regulators in different countries with overlapping supply chains.
As scanning and tracking technology becomes more widely available, the collection of additional supply-chain information presents more opportunities for identifying quality issues.
Our analysis shows the value of additional supply-chain information.

\vspace{-5pt}
\subsection{Implementation guidelines} \label{subsec:enhanceDataCollection}
\vspace{-5pt}
Implementation of the approach will be accompanied by challenges.
In addition to low overall numbers of tests, 
supply-chain information in current PMS collection records can be
limited, and this information is crucial to identifying supply chain-driven problems.
Standard PMS may benefit from supplementing the data-collection checklist proposed by \citet{newton2009}, MEDQUARG, with key supply-chain information like importers, warehouses, and intermediaries.
Furthermore, proper accounting of the uncertainty associated with a PMS test requires known sensitivity and specificity with respect to the testing tool.
Testing-tool accuracy can vary by therapeutic indication, e.g., antimalarial, or by technician experience, and thus sensitivity and specificity should be recorded for each test where possible.
In particular, false positives in low-SFP environments have the potential to confuse analysis and lead to unproductive use of resources.

The adaptable designation of nodes as individual locations or aggregates of such locations, as well as the designation of supply nodes as locations in any upstream echelon, are features that allow generalization of our approach to many\reditfin{ low-resource regulatory settings} \beditfin{low- and middle-income countries}.
The supply-chain information available to regulators is often constrained; the only requirements of our approach are standard test node labels and information, even if partial, about some upstream echelon.
Additionally, the variety in SFP causes requires adaptable methods.
For instance, economic conditions in one region may encourage a higher prevalence of falsified products in that region, or choices by one plant manager may result in a higher rate of substandard products.
Our approach allows for different analyses using individual supply-chain locations or aggregates of such locations to match goals.
The value of this adaptability is illustrated in our case study, which infers notable aggregate District SFP rates as well as SFP rates associated with individual Manufacturers.

Implementation may require customized deployments in different settings.
For instance, some settings may feature tracked as well as untracked supply-chain information; for example, scanning records may be available at every transferal point for public-sector products, while only procurement records are available for private-sector or non-profit, non-governmental products.
In this case, each test $i$ has an associated trace $\elem_i$ that is either tracked or untracked, and the vector of sourcing probabilities for untracked samples is available. 
Thus, the log-likelihood of (4),
\[\ell(\SFPrateSet|\dataSet) = \sum_{i=1}^\numTests \bigg[\log[\consolInaccFunc_{\elem_i}(\SFPrateSet)] \testResultpoint_{i} + \log[1-\consolInaccFunc_{\elem_i}(\SFPrateSet)] (1-\testResultpoint_{i})\bigg]\,,\]
can be constructed through the corresponding $\elem_i$ of each test, and inference can be conducted as described in Section 5.
Prior construction is another fundamental element of our approach involving some ambiguity.
Section \ref{subsec:casestudyanalysis} forms a prior using studies from the global literature on SFPs; using the global to characterize the local may be inadvisable in some environments.
Section \ref{sec:SFPinferenceResolutions} suggests independent priors to capture the notion that SFP generation at one node does not affect generation at other nodes.
However, it is feasible that changes in regulatory environments might stimulate correlated SFP behavior across nodes: \citet{eban2019} discussed ``two-tracked'' manufacturers with different supply lines for high-income and low-income countries.
Improved prior formation that captures local features requires an interaction between practitioners and statisticians.

In this work, the distinction between substandard and falsified is not instrumental. 
``SFP'' is broadly used to refer to products unsuitable for\reditfin{ public} consumption\reditfin{ that fail binary PMS testing}.
We consider an environment where SFPs frequently occur, test results are captured by a binary variable, and the aim is to better understand SFP sources.
Although the WHO includes unregistered products in its definition of poor quality \parencite{who2018}, our study concentrates on substandard and falsified products---the principal focus of the literature on poor-quality medical products.
Substandard and falsified products generally have different generation mechanisms \parencite{who2017-2};
however, both substandard and falsified products are problems rooted in supply-chain conditions \parencite{pisani2019}.
Usual PMS implementation seeks detection of all causes of poor quality simultaneously, and often does not require different diagnostics for each cause.
\beditfin{Regulators can select the}\reditfin{The} criteria with which tests are marked as positive or negative\reditfin{ can be set by regulator objectives}.
Depending on\reditfin{ surveillance} objectives and detection\reditfin{ capability} \beditfin{tools}, one may consider only substandard products, only falsified products, all SFPs, or even unregistered products.

The approach of this paper only considers binary pass-fail measurements, consistent with data\reditfin{ found} in MQDB.
Due to the affordability and flexibility of screening tests, PMS data can consist largely of pass-fail\reditfin{ screening} results.
Regulators can conduct a single screening test with minimal training for less than a dollar per test, while running high-powered testing requires reference standards,\reditfin{ intensive} training, and technology costing upwards of hundreds of thousands of dollars \parencite{chen2021,kovacs2014}.
Further, pharmaceuticals have different stability profiles.
Products may fail testing for any quality attribute, such as dissolution characteristics or impurity prevalence; however, proportions of expected active pharmaceutical ingredients, or API, are the most widely measured.
API content can be measured through common non-laboratory methods and\reditfin{ can} provide important information regarding\reditfin{ the} \beditfin{SFP} cause\beditfin{s}\reditfin{ of SFPs} \parencite{who2017-2}.
For instance, large discrepancies with the declared content\reditfin{ frequently} \beditfin{often} indicate falsified products.
A testing failure due to detected API content that is 4\% outside the acceptable range implies different causes than a failure with detected API content that is 70\% below the acceptable range.
The first failure is generally associated with substandard products, while the second failure is expected with falsified products.
Keeping with the binary response variable, falsified and substandard products could be categorized using different API thresholds.
However, full analysis of API\reditfin{ tests} requires modeling\reditfin{ of} the pharmaceutical-degradation process and integrating stability behavior with supply-chain information.
Moving to an inference model that treats API content as a continuous response variable would be a valuable line of future research.

\vspace{-5pt}
\subsection{Broader objectives} \label{subsec:broaderObjectives}
\vspace{-5pt}
This method can assist in the selection of testing methodology---assessing, for example, if it is better to run 1,000 tests with a spectrometer or 10,000 tests with thin-layer chromotography.
The consideration of costs versus accuracy within an inferential context could be leveraged to explore scenarios where an inexpensive, less accurate testing tool is preferential to an expensive, highly accurate testing tool, as explored in \citet{chen2021}.

The method can also inform the collection of additional samples.
If the\reditfin{ inferred} interval for a particular test node is sufficiently narrow, allocating samples to different test nodes may be recommended.
\reditfin{On the other hand }\beditfin{Alternatively}, if more data are desired regarding a particular supply node, sampling from a test node with a narrow\reditfin{ inferred} interval may be sensible if the supply node is\reditfin{ frequently} \beditfin{often} sourced by the test node.
Integration of statistical methods with regulatory insights and objectives can inform an adaptive sampling framework that feed\beditfin{s} testing results into\reditfin{ the decision of the next test node to sample} \beditfin{sample allocation decisions}.
Sequential analysis, which determines stopping rules for when data sufficient for regulator objectives have been collected, and Bayesian experiment design, which seeks to maximize the \beditfin{inference} utility\reditfin{ of inference of SFP rates} through\reditfin{ the choice of test nodes} \beditfin{sampling choices}, may be valuable avenues.
An adaptive sampling framework may also forgo the assumption that elements such as $(\SFPrateSet)$ or $\transMat$ are constant\reditfin{ over time}, and signal when these elements have significantly shifted.
Understanding how\reditfin{ a} PMS data\reditfin{ set} may be analyzed is a crucial step towards using available supply-chain information to guide the choice of sampling locations\reditfin{ that will ultimately constitute each data set}.

Additional supply-chain echelons can be integrated into the log-likelihood if supply-chain information from multiple echelons is available.
Consider a tracked case\reditfin{ with additional supply-chain information,} where each test bears a label for a node from an additional echelon of distributor nodes\reditfin{ that sits between the echelons of supplier and test nodes} \beditfin{sitting between supply nodes and test nodes}.
Let $\mathcal{C}$ be the set of distributor nodes with corresponding SFP rates $\zeta=(\zeta_1,\dots,\zeta_{|\mathcal{C}|})$.
The consolidated SFP rate of a test from a supply node $b$-distributor node $c$-test node $a$ path is then
\begin{equation}
    \consolSFPfunc_{\outletLabelpoint c\importerLabelpoint}(\outletSFPrates,\zeta, \importerSFPrates) = \outletSFPrates_{\outletLabelpoint}  + (1- \outletSFPrates_{\outletLabelpoint}) \consolSFPfunc_{c \importerLabelpoint}(\zeta,\importerSFPrates ) \,.
\end{equation}
Thus the log-likelihood of (4) can be constructed by using trace $\elem_i=(a_i,c_i,b_i)$ for each test $i$.
The consolidated SFP rate in the untracked case can be similarly formed.
Although tests with more than two labels are not explored in this paper, magnified unidentifiability issues should be anticipated when considering more than two echelons\reditfin{ for a given PMS data set}, as additional SFP rates are being inferred without additional testing data.
In a context where testing data are available from multiple echelons, future work can ascertain the conditions for identifiability or unidentifiability of SFP rates.

An additional managerial implication concerns the pooling of quality-assurance resources internationally.
The global nature of pharmaceutical supply chains means\reditfin{ SFP generation may take place at any point} \beditfin{SFPs may generate} between manufacture and domestic introduction.
Integrating\reditfin{ supply-chain} information across borders provides the possibility for studying complex, multi-tiered supply chains that feature many echelons of interconnected nodes.\reditfin{ Supply-chain information may consist of mixtures of tracked and untracked traces.}
Two countries with limited regulatory resources can expand their inferential power by sharing testing data and supply-chain information.
Expanding the scope of our approach may also entail an improved modeling of manufacturing and black-market mechanisms, perhaps using insights from\reditfin{ prior} work such as \citet{pisani2019}.
Information such as economic indicators may be incorporated into prior construction to better anticipate SFP generation. 

\if0\blind
\vspace{-10pt}
\section*{Acknowledgements}
\vspace{-5pt}
This work was funded through two National Science Foundation grants: EAGER Award 1842369: ISN: Unraveling Illicit Supply Chains for Falsified Pharmaceuticals with a Citizen Science Approach, and NSF 1953111.
We acknowledge United States Pharmacopeial Convention (USP) and the Promoting the Quality of Medicines Plus (PQM+) program, funded by the United States Agency for International Development (USAID), for the use of data in our case study and the time of staff, including the co-authors, spent contributing to this paper.
\fi

\vspace{-10pt}
\section*{Data availability statement}
\vspace{-10pt}
Due to the nature of this research, participants of this study did not agree for their data to be shared publicly; supporting data are not available. In lieu of data used in the case study, analogous synthetic data that permit similar findings can be found \if0\blind{on Github at \url{https://github.com/eugenewickett/inferringSFPsrepoducibilityreport}}\fi
\if1\blind{at [\textit{repository removed for blind review}]}\fi
.

\vspace{-10pt}
\printbibliography


\clearpage
\begin{appendices}
\section*{Appendices to ``Inferring sources of substandard and falsified products in pharmaceutical supply chains'' by Wickett, Plumlee, Smilowitz, Phanouvong and Pribluda}
\section{Proofs} \label{appdx:proofs}
\printproofs
\clearpage
\section{\beditfin{Figure for case study in the untracked setting}} \label{appdx:untrFig}
\vspace{-10pt}
\begin{figure}[h!]
    \centering
    \subfloat
    {\includegraphics[width=0.92\textwidth]{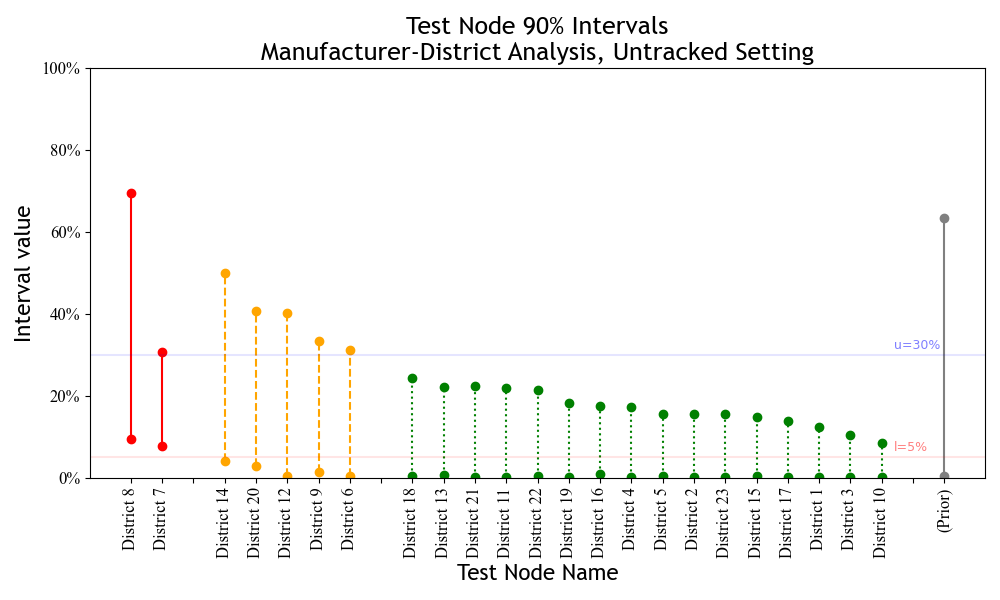} }
    \\
    \subfloat
    {\includegraphics[width=0.92\textwidth]{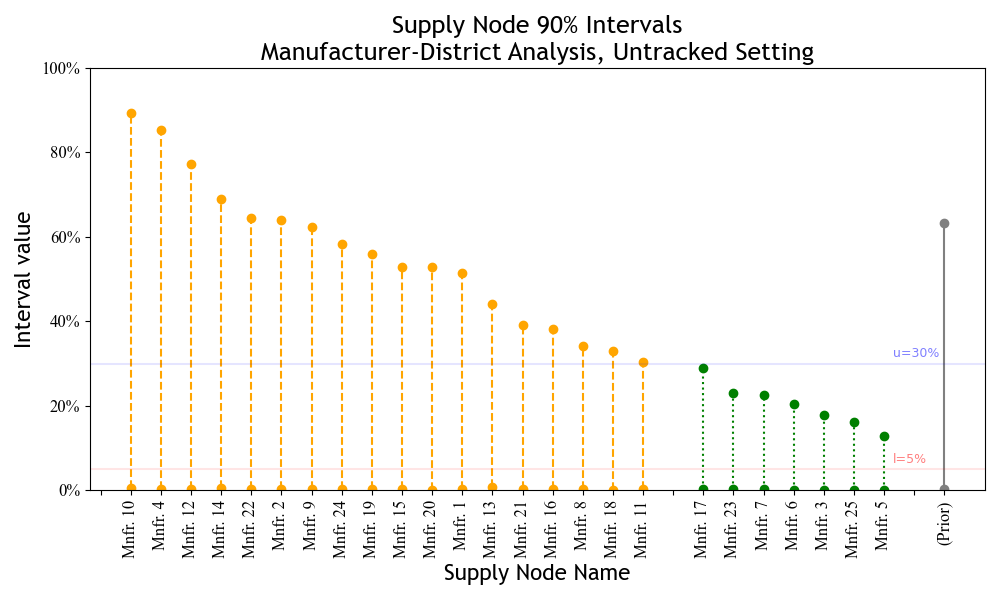}}
    \caption{\small{\beditfin{Test-node and} \reditfin{S}\beditfin{s}upply-node 90\% intervals for MCMC samples generated using case-study data \beditfin{in the untracked setting}.
    Intervals with lower bounds above $l=5\%$ are featured in solid lines on the left, intervals with lower bounds below $l$ and upper bounds above $u=30\%$ are featured in dashed lines in the middle, and all other intervals are featured in dotted lines on the right.
    }} \label{fig:manudist_SN}
\end{figure}

\end{appendices}

\clearpage
\appendix
\section*{Supplemental Online Materials to ``Inferring sources of substandard and falsified products in pharmaceutical supply chains'' by Wickett, Plumlee, Smilowitz, Phanouvong and Pribluda}
\renewcommand{\thesection}{\Roman{section}}
\section{List of log-posterior derivatives} \label{apndx:eqns}

This supplemental section lists the derivatives of the tracked and untracked log-posteriors for use with MCMC sampling.
For notation, refer to Sections \ref{sec:dataDescription} and \ref{sec:inferenceProblems} of the main document.

Let $C_p$ be the log-prior term, e.g., $C_p=\sum_{\outletLabelpoint\in \outletSet}{-\frac{1}{\expitConstantVar}|\logitfunc(\outletSFPrates_\outletLabelpoint)-\expitConstantMean|} +  \sum_{\importerLabelpoint\in \importerSet}{-\frac{1}{\expitConstantVar}|\logitfunc(\importerSFPrates_\importerLabelpoint)-\expitConstantMean| }$ in the case of the Laplace prior described in Section \ref{subsec:priorFormation}, with corresponding respective first and second log-prior derivatives of $C_1$ and $C_2$.
For the tracked case, let $\testResultpoint_{\outletLabelpoint\importerLabelpoint}=\sum_{i=1}^\numTests {\mathbbm{1}(\testResultpoint_i=1,\outletLabelpoint_i=\outletLabelpoint,\importerLabelpoint_i=\importerLabelpoint)}$ and $\numTests_{\outletLabelpoint\importerLabelpoint}=\sum_{i=1}^\numTests {\mathbbm{1}(\outletLabelpoint_i=\outletLabelpoint,\importerLabelpoint_i=\importerLabelpoint)}$, and for the untracked case let $\testResultpoint_{\outletLabelpoint}=\sum_{i=1}^\numTests {\mathbbm{1}(\testResultpoint_i=1,\outletLabelpoint_i=\outletLabelpoint)}$ and $\numTests_{\outletLabelpoint}=\sum_{i=1}^\numTests {\mathbbm{1}(\testResultpoint_i=1,\outletLabelpoint_i=\outletLabelpoint)}$.
Additionally, let $\consolInaccFunc_{\outletLabelpoint\importerLabelpoint}=\diagSens\consolSFPfunc_{\outletLabelpoint\importerLabelpoint}(\outletSFPrates,\importerSFPrates)+(1-\diagSpec)(1-\consolSFPfunc_{\outletLabelpoint\importerLabelpoint}(\outletSFPrates,\importerSFPrates))$ and
$\consolInaccFunc_{\outletLabelpoint}=\diagSens\consolSFPfunc_{\outletLabelpoint}(\outletSFPrates,\importerSFPrates)+(1-\diagSpec)(1-\consolSFPfunc_{\outletLabelpoint}(\outletSFPrates,\importerSFPrates))$ for the tracked and untracked cases, respectively:

\subsection*{Tracked case}
\subsubsection*{First derivatives}

Non-transformed test-node and supply-node SFP rates, $\outletSFPrates_\outletLabelpoint$ and $\importerSFPrates_\importerLabelpoint$, for any $\outletLabelpoint\in\outletSet$, $\importerLabelpoint\in\importerSet$:

\[\frac{\partial}{\partial \outletSFPrates_\outletLabelpoint}\log p(\SFPrateSet | \testResultpoint_1,\ldots,\testResultpoint_n) =  \sum_{\importerLabelpoint\in\importerSet}{(\diagSens+\diagSpec-1)(1-\importerSFPrates_\importerLabelpoint) \left(\frac{\testResultpoint_{\outletLabelpoint\importerLabelpoint}}{\consolInaccFunc_{\outletLabelpoint\importerLabelpoint}} - \frac{\numTests_{\outletLabelpoint\importerLabelpoint} -\testResultpoint_{\outletLabelpoint\importerLabelpoint}}{1-\consolInaccFunc_{\outletLabelpoint\importerLabelpoint}} \right)}+C_1\]

\[\frac{\partial}{\partial \importerSFPrates_\importerLabelpoint}\log p(\SFPrateSet | \testResultpoint_1,\ldots,\testResultpoint_n) =  \sum_{\outletLabelpoint\in\outletSet}{(\diagSens+\diagSpec-1)(1-\outletSFPrates_\outletLabelpoint) \left(\frac{\testResultpoint_{\outletLabelpoint\importerLabelpoint}}{\consolInaccFunc_{\outletLabelpoint\importerLabelpoint}} - \frac{\numTests_{\outletLabelpoint\importerLabelpoint} -\testResultpoint_{\outletLabelpoint\importerLabelpoint}}{1-\consolInaccFunc_{\outletLabelpoint\importerLabelpoint}} \right)}+C_1\]

\noindent
Logit-transformed test-node and supply-node SFP rates, $\outletExpit_\outletLabelpoint=\logitfunc(\outletSFPrates_\outletLabelpoint)$ and $\importerExpit_\importerLabelpoint=\logitfunc(\importerSFPrates_\importerLabelpoint)$, for any $\outletLabelpoint\in\outletSet$, $\importerLabelpoint\in\importerSet$:

\[\frac{\partial}{\partial \outletExpit_\outletLabelpoint}\log p(\SFPrateSet | \testResultpoint_1,\ldots,\testResultpoint_n) =  \sum_{\importerLabelpoint\in\importerSet}{(\diagSens+\diagSpec-1)(1-\importerSFPrates_\importerLabelpoint)(\outletSFPrates_\outletLabelpoint-\outletSFPrates_\outletLabelpoint^2)  \left(\frac{\testResultpoint_{\outletLabelpoint\importerLabelpoint}}{\consolInaccFunc_{\outletLabelpoint\importerLabelpoint}} - \frac{\numTests_{\outletLabelpoint\importerLabelpoint} -\testResultpoint_{\outletLabelpoint\importerLabelpoint}}{1-\consolInaccFunc_{\outletLabelpoint\importerLabelpoint}} \right)}+C_1\]

\[\frac{\partial}{\partial \importerExpit_\outletLabelpoint}\log p(\SFPrateSet | \testResultpoint_1,\ldots,\testResultpoint_n) =  \sum_{\outletLabelpoint\in\outletSet}{(\diagSens+\diagSpec-1)(1-\outletSFPrates_\outletLabelpoint)(\importerSFPrates_\importerLabelpoint-\importerSFPrates_\importerLabelpoint^2) \left(\frac{\testResultpoint_{\outletLabelpoint\importerLabelpoint}}{\consolInaccFunc_{\outletLabelpoint\importerLabelpoint}} - \frac{\numTests_{\outletLabelpoint\importerLabelpoint} -\testResultpoint_{\outletLabelpoint\importerLabelpoint}}{1-\consolInaccFunc_{\outletLabelpoint\importerLabelpoint}} \right)}+C_1\]

\subsubsection*{Second derivatives}
Non-transformed test-node and supply-node SFP rates, $\outletSFPrates_\outletLabelpoint$ and $\importerSFPrates_\importerLabelpoint$, for any $\outletLabelpoint\in\outletSet$, $\importerLabelpoint\in\importerSet$:

\[\frac{\partial^2}{\partial\outletSFPrates_\outletLabelpoint^2}\log p(\SFPrateSet | \testResultpoint_1,\ldots,\testResultpoint_n) =  \sum_{\importerLabelpoint\in\importerSet}{(\diagSens+\diagSpec-1)^2(1-\importerSFPrates_\importerLabelpoint)^2 \left(-\frac{\testResultpoint_{\outletLabelpoint\importerLabelpoint}}{\consolInaccFunc_{\outletLabelpoint\importerLabelpoint}^2} - \frac{\numTests_{\outletLabelpoint\importerLabelpoint} -\testResultpoint_{\outletLabelpoint\importerLabelpoint}}{(1-\consolInaccFunc_{\outletLabelpoint\importerLabelpoint})^2} \right)}+\priorHessConst\]

\[\frac{\partial^2}{\partial \importerSFPrates_\importerLabelpoint^2}\log p(\SFPrateSet | \testResultpoint_1,\ldots,\testResultpoint_n) =  \sum_{\outletLabelpoint\in\outletSet}{(\diagSens+\diagSpec-1)^2(1-\outletSFPrates_\outletLabelpoint)^2 \left(-\frac{\testResultpoint_{\outletLabelpoint\importerLabelpoint}}{\consolInaccFunc_{\outletLabelpoint\importerLabelpoint}^2} - \frac{\numTests_{\outletLabelpoint\importerLabelpoint} -\testResultpoint_{\outletLabelpoint\importerLabelpoint}}{(1-\consolInaccFunc_{\outletLabelpoint\importerLabelpoint})^2} \right)}+\priorHessConst\]

\begin{align*}
\frac{\partial^2}{\partial\outletSFPrates_\outletLabelpoint\importerSFPrates_\importerLabelpoint}\log p(\SFPrateSet | \testResultpoint_1,\ldots,\testResultpoint_n) &=  (\diagSens+\diagSpec-1)^2(1-\importerSFPrates_\importerLabelpoint)(1-\outletSFPrates_\outletLabelpoint) \left(-\frac{\testResultpoint_{\outletLabelpoint\importerLabelpoint}}{\consolInaccFunc_{\outletLabelpoint\importerLabelpoint}^2} - \frac{\numTests_{\outletLabelpoint\importerLabelpoint} -\testResultpoint_{\outletLabelpoint\importerLabelpoint}}{(1-\consolInaccFunc_{\outletLabelpoint\importerLabelpoint})^2} \right)\\
&-(\diagSens+\diagSpec-1) \left( \frac{\testResultpoint_{\outletLabelpoint\importerLabelpoint}}{\consolInaccFunc_{\outletLabelpoint\importerLabelpoint}} - \frac{\numTests_{\outletLabelpoint\importerLabelpoint} -\testResultpoint_{\outletLabelpoint\importerLabelpoint}}{1-\consolInaccFunc_{\outletLabelpoint\importerLabelpoint}}  \right)+\priorHessConst
\end{align*}

\noindent
Logit-transformed test-node and supply-node SFP rates, $\outletExpit_\outletLabelpoint$ and $\importerExpit_\importerLabelpoint$, for any $\outletLabelpoint\in\outletSet$, $\importerLabelpoint\in\importerSet$:

\begin{align*}
\frac{\partial^2}{\partial\outletExpit_\outletLabelpoint^2}\log p(\SFPrateSet | \testResultpoint_1,\ldots,\testResultpoint_n) &=
\sum_{\importerLabelpoint\in\importerSet}{(\diagSens+\diagSpec-1)^2(1-\importerSFPrates_\importerLabelpoint)^2(\outletSFPrates_\outletLabelpoint-\outletSFPrates_\outletLabelpoint^2)^2 \left(-\frac{\testResultpoint_{\outletLabelpoint\importerLabelpoint}}{\consolInaccFunc_{\outletLabelpoint\importerLabelpoint}^2} - \frac{\numTests_{\outletLabelpoint\importerLabelpoint} -\testResultpoint_{\outletLabelpoint\importerLabelpoint}}
{(1-\consolInaccFunc_{\outletLabelpoint\importerLabelpoint})^2} \right)}\\
&+ (\diagSens+\diagSpec-1)(1-\importerSFPrates_\importerLabelpoint)(\outletSFPrates_\outletLabelpoint-3\outletSFPrates_\outletLabelpoint^2+2\outletSFPrates_\outletLabelpoint^3)
\left(\frac{\testResultpoint_{\outletLabelpoint\importerLabelpoint}}{\consolInaccFunc_{\outletLabelpoint\importerLabelpoint}} - \frac{\numTests_{\outletLabelpoint\importerLabelpoint} -\testResultpoint_{\outletLabelpoint\importerLabelpoint}}
{1-\consolInaccFunc_{\outletLabelpoint\importerLabelpoint}} \right)+\priorHessConst
\end{align*}

\begin{align*}
\frac{\partial^2}{\partial\importerExpit_\importerLabelpoint^2}\log p(\SFPrateSet | \testResultpoint_1,\ldots,\testResultpoint_n) &=
\sum_{\outletLabelpoint\in\outletSet}{(\diagSens+\diagSpec-1)^2(1-\outletSFPrates_\outletLabelpoint)^2(\importerSFPrates_\importerLabelpoint-\importerSFPrates_\importerLabelpoint^2)^2 \left(-\frac{\testResultpoint_{\outletLabelpoint\importerLabelpoint}}{\consolInaccFunc_{\outletLabelpoint\importerLabelpoint}^2} - \frac{\numTests_{\outletLabelpoint\importerLabelpoint} -\testResultpoint_{\outletLabelpoint\importerLabelpoint}}
{(1-\consolInaccFunc_{\outletLabelpoint\importerLabelpoint})^2} \right)}\\
&+ (\diagSens+\diagSpec-1)(1-\outletSFPrates_\outletLabelpoint)(\importerSFPrates_\importerLabelpoint-3\importerSFPrates_\importerLabelpoint^2+2\importerSFPrates_\importerLabelpoint^3)
\left(\frac{\testResultpoint_{\outletLabelpoint\importerLabelpoint}}{\consolInaccFunc_{\outletLabelpoint\importerLabelpoint}} - \frac{\numTests_{\outletLabelpoint\importerLabelpoint} -\testResultpoint_{\outletLabelpoint\importerLabelpoint}}
{1-\consolInaccFunc_{\outletLabelpoint\importerLabelpoint}} \right)+\priorHessConst
\end{align*}

\begin{align*}
\frac{\partial^2}{\partial\outletExpit_\outletLabelpoint\importerExpit_\importerLabelpoint}\log p(\SFPrateSet | \testResultpoint_1,\ldots,\testResultpoint_n) &=
(\diagSens+\diagSpec-1)^2(1-\outletSFPrates_\outletLabelpoint)^2(1-\importerSFPrates_\importerLabelpoint^2)^2\outletSFPrates_\outletLabelpoint\importerSFPrates_\importerLabelpoint \left(-\frac{\testResultpoint_{\outletLabelpoint\importerLabelpoint}}{\consolInaccFunc_{\outletLabelpoint\importerLabelpoint}^2} - \frac{\numTests_{\outletLabelpoint\importerLabelpoint} -\testResultpoint_{\outletLabelpoint\importerLabelpoint}}
{(1-\consolInaccFunc_{\outletLabelpoint\importerLabelpoint})^2} \right)\\
&- (\diagSens+\diagSpec-1)(1-\outletSFPrates_\outletLabelpoint)(1-\importerSFPrates_\importerLabelpoint)\outletSFPrates_\outletLabelpoint\importerSFPrates_\importerLabelpoint
\left(\frac{\testResultpoint_{\outletLabelpoint\importerLabelpoint}}{\consolInaccFunc_{\outletLabelpoint\importerLabelpoint}} - \frac{\numTests_{\outletLabelpoint\importerLabelpoint} -\testResultpoint_{\outletLabelpoint\importerLabelpoint}}
{1-\consolInaccFunc_{\outletLabelpoint\importerLabelpoint}} \right)+\priorHessConst
\end{align*}

\subsection*{Untracked case}
\subsubsection*{First derivatives}

Non-transformed test-node and supply-node SFP rates, $\outletSFPrates_\outletLabelpoint$ and $\importerSFPrates_\importerLabelpoint$, for any $\outletLabelpoint\in\outletSet$, $\importerLabelpoint\in\importerSet$:

\[\frac{\partial}{\partial \outletSFPrates_\outletLabelpoint}\log p(\SFPrateSet | \testResultpoint_1,\ldots,\testResultpoint_n) =  (\diagSens+\diagSpec-1)(1-\transMat_{\outletLabelpoint}\importerSFPrates) \left(\frac{\testResultpoint_{\outletLabelpoint}}{\consolInaccFunc_{\outletLabelpoint}} - \frac{\numTests_{\outletLabelpoint} -\testResultpoint_{\outletLabelpoint}}{1-\consolInaccFunc_{\outletLabelpoint}} \right)+C_1\]

\[\frac{\partial}{\partial \importerSFPrates_\importerLabelpoint}\log p(\SFPrateSet | \testResultpoint_1,\ldots,\testResultpoint_n) =  \sum_{\outletLabelpoint\in\outletSet}{(\diagSens+\diagSpec-1)(1-\outletSFPrates_\outletLabelpoint)\transMat_{\outletLabelpoint\importerLabelpoint} \left(\frac{\testResultpoint_{\outletLabelpoint}}{\consolInaccFunc_{\outletLabelpoint}} - \frac{\numTests_{\outletLabelpoint} -\testResultpoint_{\outletLabelpoint}}{1-\consolInaccFunc_{\outletLabelpoint}} \right)}+C_1\]

\noindent
Logit-transformed test-node and supply-node SFP rates, $\outletExpit_\outletLabelpoint$ and $\importerExpit_\importerLabelpoint$, for any $\outletLabelpoint\in\outletSet$, $\importerLabelpoint\in\importerSet$:

\[\frac{\partial}{\partial \outletExpit_\outletLabelpoint}\log p(\SFPrateSet | \testResultpoint_1,\ldots,\testResultpoint_n) =  (\diagSens+\diagSpec-1)(1-\transMat_{\outletLabelpoint}\importerSFPrates)(\outletSFPrates_\outletLabelpoint-\outletSFPrates_\outletLabelpoint^2) \left(\frac{\testResultpoint_{\outletLabelpoint}}{\consolInaccFunc_{\outletLabelpoint}} - \frac{\numTests_{\outletLabelpoint} -\testResultpoint_{\outletLabelpoint}}{1-\consolInaccFunc_{\outletLabelpoint}} \right)+C_1\]

\[\frac{\partial}{\partial \importerExpit_\outletLabelpoint}\log p(\SFPrateSet | \testResultpoint_1,\ldots,\testResultpoint_n) =  \sum_{\outletLabelpoint\in\outletSet}{(\diagSens+\diagSpec-1)(1-\outletSFPrates_\outletLabelpoint)(\importerSFPrates_\importerLabelpoint-\importerSFPrates_\importerLabelpoint^2)\transMat_{\outletLabelpoint\importerLabelpoint} \left(\frac{\testResultpoint_{\outletLabelpoint}}{\consolInaccFunc_{\outletLabelpoint}} - \frac{\numTests_{\outletLabelpoint} -\testResultpoint_{\outletLabelpoint}}{1-\consolInaccFunc_{\outletLabelpoint}} \right)}+C_1\]

\subsubsection*{Second derivatives}
Non-transformed test-node and supply-node SFP rates, $\outletSFPrates_\outletLabelpoint$ and $\importerSFPrates_\importerLabelpoint$, for any $\outletLabelpoint\in\outletSet$, $\importerLabelpoint\in\importerSet$, with $\importerLabelpoint'\neq\importerLabelpoint$:

\[\frac{\partial^2}{\partial \outletSFPrates_\outletLabelpoint^2}\log p(\SFPrateSet | \testResultpoint_1,\ldots,\testResultpoint_n) =  (\diagSens+\diagSpec-1)^2(1-\transMat_{\outletLabelpoint}\importerSFPrates)^2 \left(-\frac{\testResultpoint_{\outletLabelpoint}}{\consolInaccFunc_{\outletLabelpoint}^2} - \frac{\numTests_{\outletLabelpoint} -\testResultpoint_{\outletLabelpoint}}{(1-\consolInaccFunc_{\outletLabelpoint})^2} \right)+\priorHessConst\]

\[\frac{\partial^2}{\partial \importerSFPrates_\importerLabelpoint^2}\log p(\SFPrateSet | \testResultpoint_1,\ldots,\testResultpoint_n) =  \sum_{\outletLabelpoint\in\outletSet}{(\diagSens+\diagSpec-1)^2(1-\outletSFPrates_\outletLabelpoint)^2\transMat_{\outletLabelpoint\importerLabelpoint}^2 \left(-\frac{\testResultpoint_{\outletLabelpoint}}{\consolInaccFunc_{\outletLabelpoint}^2} - \frac{\numTests_{\outletLabelpoint} -\testResultpoint_{\outletLabelpoint}}{(1-\consolInaccFunc_{\outletLabelpoint})^2} \right)}+\priorHessConst\]

\begin{align*}
\frac{\partial^2}{\partial\outletSFPrates_\outletLabelpoint\importerSFPrates_\importerLabelpoint}\log p(\SFPrateSet | \testResultpoint_1,\ldots,\testResultpoint_n) &=  \transMat_{\outletLabelpoint\importerLabelpoint}(\diagSens+\diagSpec-1)^2(1-\outletSFPrates_\outletLabelpoint)(1-\transMat_{\outletLabelpoint}\importerSFPrates) \left(-\frac{\testResultpoint_{\outletLabelpoint}}{\consolInaccFunc_{\outletLabelpoint}^2} - \frac{\numTests_{\outletLabelpoint} -\testResultpoint_{\outletLabelpoint}}{(1-\consolInaccFunc_{\outletLabelpoint})^2} \right)\\
&-\transMat_{\outletLabelpoint\importerLabelpoint}(\diagSens+\diagSpec-1)\left(\frac{\testResultpoint_{\outletLabelpoint}}{\consolInaccFunc_{\outletLabelpoint}} - \frac{\numTests_{\outletLabelpoint} -\testResultpoint_{\outletLabelpoint}}{1-\consolInaccFunc_{\outletLabelpoint}}\right)+\priorHessConst
\end{align*}

\[\frac{\partial^2}{\partial \importerSFPrates_\importerLabelpoint\importerSFPrates_{\importerLabelpoint'}}\log p(\SFPrateSet | \testResultpoint_1,\ldots,\testResultpoint_n) =  \sum_{\outletLabelpoint\in\outletSet}{(\diagSens+\diagSpec-1)^2(1-\outletSFPrates_\outletLabelpoint)^2\transMat_{\outletLabelpoint\importerLabelpoint}\transMat_{\outletLabelpoint\importerLabelpoint'} \left(-\frac{\testResultpoint_{\outletLabelpoint}}{\consolInaccFunc_{\outletLabelpoint}^2} - \frac{\numTests_{\outletLabelpoint} -\testResultpoint_{\outletLabelpoint}}{(1-\consolInaccFunc_{\outletLabelpoint})^2} \right)}+\priorHessConst\]

\noindent
Logit-transformed test-node and supply-node SFP rates, $\outletExpit=\expitfunc(\outletSFPrates)$ and $\importerExpit=\expitfunc(\importerSFPrates)$, for any $\outletLabelpoint\in\outletSet$, $\importerLabelpoint\in\importerSet$, with $\importerLabelpoint'\neq\importerLabelpoint$:

\begin{align*}
\frac{\partial^2}{\partial \outletExpit_\outletLabelpoint^2}\log p(\SFPrateSet | \testResultpoint_1,\ldots,\testResultpoint_n) &=  (\diagSens+\diagSpec-1)^2(1-\transMat_{\outletLabelpoint}\importerSFPrates)^2(\outletSFPrates_\outletLabelpoint-\outletSFPrates_\outletLabelpoint^2)^2 \left(-\frac{\testResultpoint_{\outletLabelpoint}}{\consolInaccFunc_{\outletLabelpoint}^2} - \frac{\numTests_{\outletLabelpoint} -\testResultpoint_{\outletLabelpoint}}{(1-\consolInaccFunc_{\outletLabelpoint})^2} \right)\\
&+(\diagSens+\diagSpec-1)(1-\transMat_{\outletLabelpoint}\importerSFPrates)(\outletSFPrates_\outletLabelpoint-3\outletSFPrates_\outletLabelpoint^2+2\outletSFPrates_\outletLabelpoint^3) \left(\frac{\testResultpoint_{\outletLabelpoint}}{\consolInaccFunc_{\outletLabelpoint}} - \frac{\numTests_{\outletLabelpoint} -\testResultpoint_{\outletLabelpoint}}{1-\consolInaccFunc_{\outletLabelpoint}} \right)+\priorHessConst
\end{align*}

\begin{align*}
\frac{\partial^2}{\partial \importerExpit_\importerLabelpoint^2}\log p(\SFPrateSet | \testResultpoint_1,\ldots,\testResultpoint_n) &=  \sum_{\outletLabelpoint\in\outletSet}{(\diagSens+\diagSpec-1)^2(1-\outletSFPrates_\outletLabelpoint)^2(\importerSFPrates_\importerLabelpoint-\importerSFPrates_\importerLabelpoint^2)^2\transMat_{\outletLabelpoint\importerLabelpoint}^2 \left(-\frac{\testResultpoint_{\outletLabelpoint}}{\consolInaccFunc_{\outletLabelpoint}^2} - \frac{\numTests_{\outletLabelpoint} -\testResultpoint_{\outletLabelpoint}}{(1-\consolInaccFunc_{\outletLabelpoint})^2} \right)}\\
&+(\diagSens+\diagSpec-1)(1-\outletSFPrates_\outletLabelpoint)(\importerSFPrates_\importerLabelpoint-3\importerSFPrates_\importerLabelpoint^2+2\importerSFPrates_\importerLabelpoint^3)\transMat_{\outletLabelpoint\importerLabelpoint} \left(\frac{\testResultpoint_{\outletLabelpoint}}{\consolInaccFunc_{\outletLabelpoint}} - \frac{\numTests_{\outletLabelpoint} -\testResultpoint_{\outletLabelpoint}}{1-\consolInaccFunc_{\outletLabelpoint}} \right)+\priorHessConst
\end{align*}

\begin{align*}
\frac{\partial^2}{\partial\outletExpit_\outletLabelpoint\importerExpit_\importerLabelpoint}\log p(\SFPrateSet | \testResultpoint_1,\ldots,\testResultpoint_n) &=  \transMat_{\outletLabelpoint\importerLabelpoint}(\diagSens+\diagSpec-1)^2(1-\outletSFPrates_\outletLabelpoint)(1-\transMat_{\outletLabelpoint}\importerSFPrates)(\outletSFPrates_\outletLabelpoint-\outletSFPrates_\outletLabelpoint^2)(\importerSFPrates_\importerLabelpoint-\importerSFPrates_\importerLabelpoint^2) \left(-\frac{\testResultpoint_{\outletLabelpoint}}{\consolInaccFunc_{\outletLabelpoint}^2} - \frac{\numTests_{\outletLabelpoint} -\testResultpoint_{\outletLabelpoint}}{(1-\consolInaccFunc_{\outletLabelpoint})^2} \right)\\
&-\transMat_{\outletLabelpoint\importerLabelpoint}(\diagSens+\diagSpec-1)(\outletSFPrates_\outletLabelpoint-\outletSFPrates_\outletLabelpoint^2)(\importerSFPrates_\importerLabelpoint-\importerSFPrates_\importerLabelpoint^2)
\left(\frac{\testResultpoint_{\outletLabelpoint}}{\consolInaccFunc_{\outletLabelpoint}} - \frac{\numTests_{\outletLabelpoint} -\testResultpoint_{\outletLabelpoint}}{1-\consolInaccFunc_{\outletLabelpoint}}\right)+\priorHessConst
\end{align*}

\[\frac{\partial^2}{\partial \importerExpit_\importerLabelpoint\importerExpit_{\importerLabelpoint'}}\log p(\SFPrateSet | \testResultpoint_1,\ldots,\testResultpoint_n) =  \sum_{\outletLabelpoint\in\outletSet}{\transMat_{\outletLabelpoint\importerLabelpoint}\transMat_{\outletLabelpoint\importerLabelpoint'}(\diagSens+\diagSpec-1)^2(1-\outletSFPrates_\outletLabelpoint)^2(\importerSFPrates_\importerLabelpoint-\importerSFPrates_\importerLabelpoint^2)(\importerSFPrates_{\importerLabelpoint'}-(\importerSFPrates_{\importerLabelpoint'})^2) \left(-\frac{\testResultpoint_{\outletLabelpoint}}{\consolInaccFunc_{\outletLabelpoint}^2} - \frac{\numTests_{\outletLabelpoint} -\testResultpoint_{\outletLabelpoint}}{(1-\consolInaccFunc_{\outletLabelpoint})^2} \right)}+\priorHessConst\]
\section{MCMC computation} \label{apndx:mcmcDetails}

This supplemental section discusses computational considerations for our methods.
We first use synthetic data to explore computation time and the choice of the Markov chain Monte Carlo (MCMC) sampler.
Table \ref{tab:computationtime} shows computation time statistics for MCMC samples drawn using synthetic data in different scenarios for twenty different supply chains.
In each scenario, 1000 tests are uniformly sampled across all test nodes.
The vector of sourcing probabilities for each test node is randomly generated using a Pareto distribution to capture real-world sourcing behavior.
Increasing or decreasing the Pareto scale parameter increases or decreases the trace density.
Trace density refers to the proportion of possible supply-chain traces observed in the data set.
For example, if 22 unique supply-chain traces are observed for a system with 10 test nodes and 10 supply nodes, the trace density is $\frac{22}{10\times 10}=0.22$.
We consider the Langevin Monte Carlo (LMC) sampler of \citet{surmise2021} in addition to the No-U-Turn Sampler (NUTS) sampler of \citet{hoffman2014no}.
MCMC samples are drawn using either NUTS or LMC on an Intel Core 1.6GHz processor.
Trace density has a small but negligible effect on computation time: doubling the trace density from 0.079 to 0.169 resulted in a three-second decrease in average computation time (about 15\%).
Having more supply-chain traces narrows the region of SFP rates that credibly explain the data, thus finding suitable draws is easier for the MCMC sampler.
The size of the supply chain has a significant effect on computation time: increasing the total number of nodes from fifty to two hundred increases the average computation time by a factor of three.

\begin{table}[b]
\makegapedcells
\centering
\caption{\textit{Computation time statistics for MCMC-sample generation under select scenarios.
}}
\begin{tabular}{P{0.7cm}P{0.7cm}P{2cm}P{2cm}P{2cm}P{2cm}P{2cm}}
\toprule
$|\outletSet|$ & $|\importerSet|$ & \small{MCMC sampler} & \small{Avg. trace density}  & \small{Avg. comp. time (s)} & \small{Min. comp. time (s)} & \small{Max. comp. time (s)}
\\ \hline
   50 & 50 & \small{NUTS} & 0.133 & 19.2 & 14.4 & 37.0
\\ \hline
    50 & 50 & \small{LMC} & 0.133 & 31.5 & 15.4 & 78.1
\\ \hline
    50 & 50 & \small{NUTS} & 0.079 & 21.2 & 17.1 & 39.1
\\ \hline
    50 & 50 & \small{NUTS} & 0.169 & 17.9 & 14.7 & 30.8
\\ \hline
    25 & 25 & \small{NUTS} & 0.270 & 14.8 & 10.0 & 40.6
\\ \hline
    100 & 100 & \small{NUTS} & 0.049 & 46.9 & 29.9 & 73.7
\\
\bottomrule
\end{tabular}
 \label{tab:computationtime}
\end{table}

\begin{figure}
    \centering
    \includegraphics[width=0.9\textwidth]{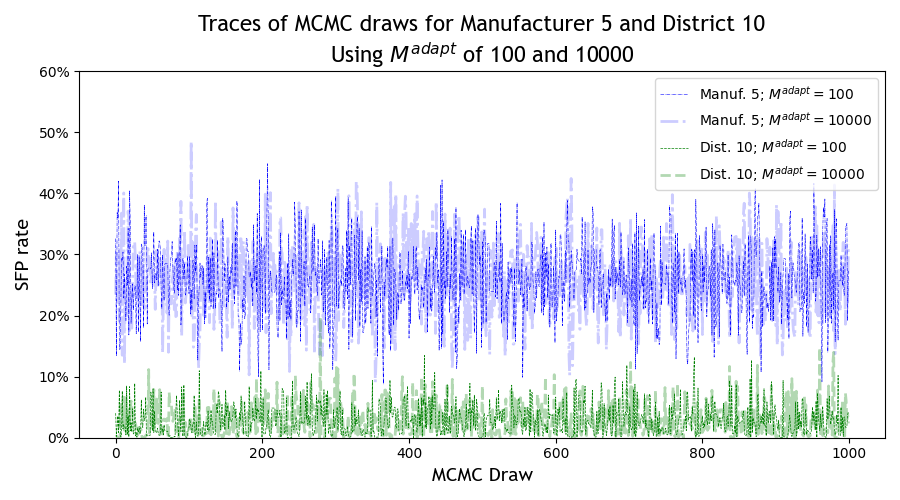}
    \caption{Trace plot for inference draws under NUTS for SFP rates at Manufacture 5 and District 10 of the case study. 
    $M^{adapt}$ is the number of warm-start draws used to determine tuning parameters for NUTS.}
    \label{fig:mcmcTrace}
\end{figure}

We next illustrate the sufficiency of using 5,000 warm-start draws for our case study data.
\citeauthor{hoffman2014no} showed that 1,000 warm-start draws were sufficient for convergence of the tuning parameters used in NUTS for some chosen target distributions.
These target distributions featured dimensionality in the hundreds or thousands as well as strong correlations.
For lower targeted acceptance rates of proposal samples, these problems required more warm-start draws for tuning parameters to sufficiently converge.
\if0\blind{Software implementation \texttt{logistigate} \parencite{logistigate2021} }\fi
\if1\blind{[\textit{Software removed for blind review}] }\fi uses a target acceptance rate of 0.4, which is the lower end of the acceptance rate threshold suggested by \citeauthor{hoffman2014no}.
Using 5,000 warm-start draws is a conservative choice for the warm-start period relative to \citeauthor{hoffman2014no}'s target distributions.
Figure \ref{fig:mcmcTrace} shows traces for the SFP rates of Manufacturer 5 and District 10 of the case study, using warm-start periods ($M^{adapt}$ in \citeauthor{hoffman2014no}) of 100 and 10,000 draws.
Visual inspection indicates that the tuning parameters generated during warm-starts of 100 and 10,000 draws produce similar distributions of the SFP rates at Manufacturer 5 and District 10. 
These nodes are chosen for illustrative purposes, but the observation holds for other nodes as well.
Thus, we find 5,000 draws to be a sufficient warm-start period.

\begin{figure}
    \centering
    \includegraphics[width=0.9\textwidth]{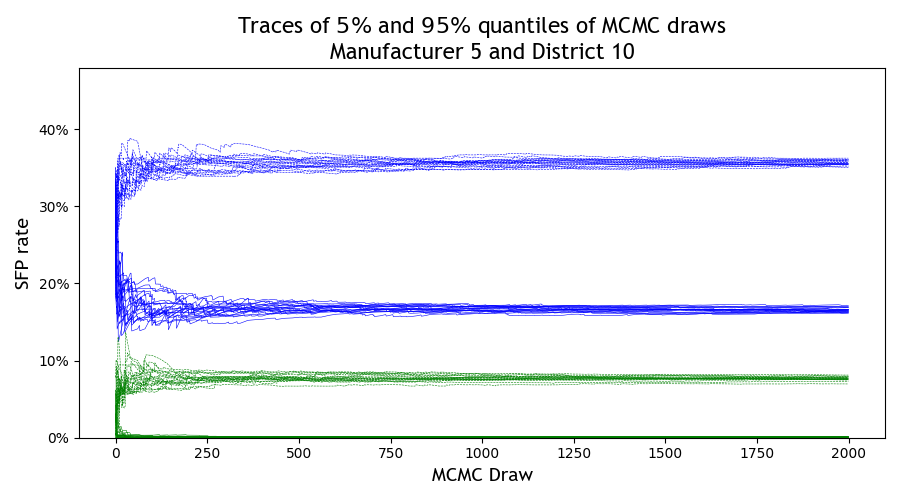}
    \caption{Plots of cumulative 5\% (solid lines) and 95\% (dashed lines) quantiles for twenty sets of MCMC draws under NUTS for SFP rates at Manufacture 5 (in blue) and District 10 (in green) of the case study.}
    \label{fig:mcmcTraceQuants}
\end{figure}

The case study uses 90\% credible intervals under 1,000 MCMC draws for inference.
Figure \ref{fig:mcmcTraceQuants} shows cumulative 5\% and 95\% quantiles as the number of inference draws increases to 2,000 for the SFP rates associated with Manufacturer 5 and District 10, for twenty separate runs using NUTS.
Visual inspection suggests that 1,000 draws for inference is sufficient for analysis in the case study.

\section{Sensitivity analysis for case study} \label{apndx:caseStudySensitivity}
We use the case-study data to analyze the impact that prior choice, testing-tool uncertainty, and estimation of the sourcing-probability matrix $\transMat$ have on inference for SFP rates associated with Manufacturers 5, 8, and 10, as well as Districts 7 and 8.
This node set contains all nodes with significantly high SFP rates as discussed in Section \ref{subsec:casestudyanalysis}.

\begin{table}
\makegapedcells
\centering
\caption{\textit{Widths for MCMC-sample 90\% intervals associated with chosen Districts and Manufacturers under different values for testing sensitivity, $\diagSens$, and specificity, $\diagSpec$.
For example, an interval of $[42\%, 55\%]$ has a width of 13\%.
$n$ denotes the number of samples featuring each node, and $y$ denotes the number of corresponding positive SFP tests.
}} \label{tab:toolsensitivity}
\begin{tabular}{P{0.7cm}P{0.7cm}P{1.9cm}P{1.9cm}P{1.9cm}P{1.9cm}P{1.9cm}}
\toprule
$\diagSens$ & $\diagSpec$ & \makecell[t]{\rot{90}{\footnotesize{\textbf{Mnfr. 5}}} \\ \scriptsize{$n=82$} \\ \scriptsize{$y=28$}} & \makecell[t]{\rot{90}{\footnotesize{\textbf{Mnfr. 8}}} \\ \scriptsize{$n=21$} \\ \scriptsize{$y=8$}} & \makecell[t]{\rot{90}{\footnotesize{\textbf{Mnfr. 10}}} \\ \scriptsize{$n=7$} \\ \scriptsize{$y=4$}} & \makecell[t]{\rot{90}{\footnotesize{\textbf{District 7}}} \\ \scriptsize{$n=81$} \\ \scriptsize{$y=24$}} & \makecell[t]{\rot{90}{\footnotesize{\textbf{District 8}}} \\ \scriptsize{$n=12$} \\ \scriptsize{$y=7$}}
\\ \hline
   1.0 & 1.0 & 19.3\% & 35.4\% & 65.5\% & 18.5\% & 58.3\%
\\ \hline
   0.8 & 1.0 & 24.8\% & 47.7\% & 80.6\% & 24.1\% & 69.6\% 
\\ \hline
   1.0 & 0.95 & 20.0\% & 36.7\% & 62.5\% & 18.2\% & 58.6\%
\\ \hline
   0.8 & 0.95 & 25.8\% & 49.4\% & 84.5\% & 26.2\% & 75.6\%
\\
\bottomrule
\end{tabular}
\end{table}

We demonstrate the impact of reduced testing-tool accuracy by alternately assuming a sensitivity of $\diagSens=80\%$ and a specificity of $\diagSpec=95\%$ under the same Manufacturer-District analysis.
These sensitivity and specificity values fall within the ranges described for different testing tools in \citet{kovacs2014}.
Table \ref{tab:toolsensitivity} illustrates the changes in interval widths when reducing testing-tool accuracy.
$n$ under each node label denotes the number of samples associated with that node, while $y$ denotes the number of associated positive tests.
As expected, testing-tool uncertainty generally has an inflationary effect on inference, and this inflationary effect is larger for nodes for which there are less data.
The interval width for Manufacturer 5 increases by about six percentage points when $\diagSens=80\%$ and $\diagSpec=95\%$, while the interval width for Manufacturer 10 increases by nineteen percentage points. 
The possibility of false negatives and false positives creates more credible explanations for observed patterns.

\begin{table}
\makegapedcells
\centering
\caption{\textit{Widths for MCMC-sample 90\% intervals associated with chosen Districts and Manufacturers under different choices of prior.
For example, an interval of $[42\%, 55\%]$ has a width of 13\%.
$\expitConstantMean$ denotes the prior average and $\expitConstantVar$ denotes the scale parameter in the Laplace case or the variance in the normal case.
$n$ denotes the number of samples featuring each node, and $y$ denotes the number of corresponding positive SFP tests.
}} \label{tab:priorsensitivity}
\begin{tabular}{P{0.7cm}P{0.7cm}P{1.5cm}P{1.9cm}P{1.9cm}P{1.9cm}P{1.9cm}P{1.9cm}}
\toprule
$\expitConstantMean$ & $\expitConstantVar$ & \small{Laplace/ Normal} & \makecell[t]{\rot{90}{\footnotesize{\textbf{Mnfr. 5}}} \\ \scriptsize{$n=82$} \\ \scriptsize{$y=28$}} & \makecell[t]{\rot{90}{\footnotesize{\textbf{Mnfr. 8}}} \\ \scriptsize{$n=21$} \\ \scriptsize{$y=8$}} & \makecell[t]{\rot{90}{\footnotesize{\textbf{Mnfr. 10}}} \\ \scriptsize{$n=7$} \\ \scriptsize{$y=4$}} & \makecell[t]{\rot{90}{\footnotesize{\textbf{District 7}}} \\ \scriptsize{$n=81$} \\ \scriptsize{$y=24$}} & \makecell[t]{\rot{90}{\footnotesize{\textbf{District 8}}} \\ \scriptsize{$n=12$} \\ \scriptsize{$y=7$}}
\\ \hline
   -2.5 & 1.30 & \small{Laplace} & 19.3\% & 35.4\% & 65.5\% & 18.5\% & 58.3\%
\\ \hline
   -3.5 & 1.30 & \small{Laplace} & 18.5\% & 35.4\% & 64.1\% & 18.0\% & 63.8\% 
\\ \hline
   -2.5 & 0.87 & \small{Laplace} & 19.4\% & 36.2\% & 58.7\% & 18.3\% & 55.8\%
\\ \hline
   -2.5 & 3.38 & \small{Normal} & 19.3\% & 34.1\% & 67.6\% & 16.4\% & 54.3\%
\\
\bottomrule
\end{tabular}
\end{table}

Table \ref{tab:priorsensitivity} displays changes in 90\% interval widths under different choices for the prior.
In addition to the prior introduced in Section \ref{subsec:casestudyanalysis} with average $\expitConstantMean=-2.5$ and spread parameter $\expitConstantVar=1.3$ (associated 90\% interval of [0.4\%, 63\%]), the table also depicts a prior with a reduced average of $\expitConstantMean=-3.5$ ([0.1\%, 39\%]), and a prior with a reduced spread of $0.87$ ([1\%, 38\%]).
The table also shows results under an independent normal prior with average $\expitConstantMean=-2.5$ and spread parameter $\expitConstantVar=3.38$ ([0.4\%, 64\%]), which has an identical average and variance to an independent Laplace with average $\expitConstantMean=-2.5$ and spread $\expitConstantVar=1.3$.
A reduction in prior average implies a lower anticipation of the overall SFP rate, while a reduction in prior spread suggests higher confidence that node SFP rates will be similar.
The table indicates that prior choice from within this set of priors does not have an instrumental effect on interval width;
sufficient data seem to overwhelm the prior designation.
When data associated with a node are small, as in the case of Manufacturer 10, interval widths vary as much as ten percentage points.
As data associated with a node increase, the corresponding interval widths only vary by one or two percentage points across these different choices of prior, as in the case of Manufacturer 5.

\begin{figure}
    \centering
    \subfloat
    {\includegraphics[width=0.95\textwidth]{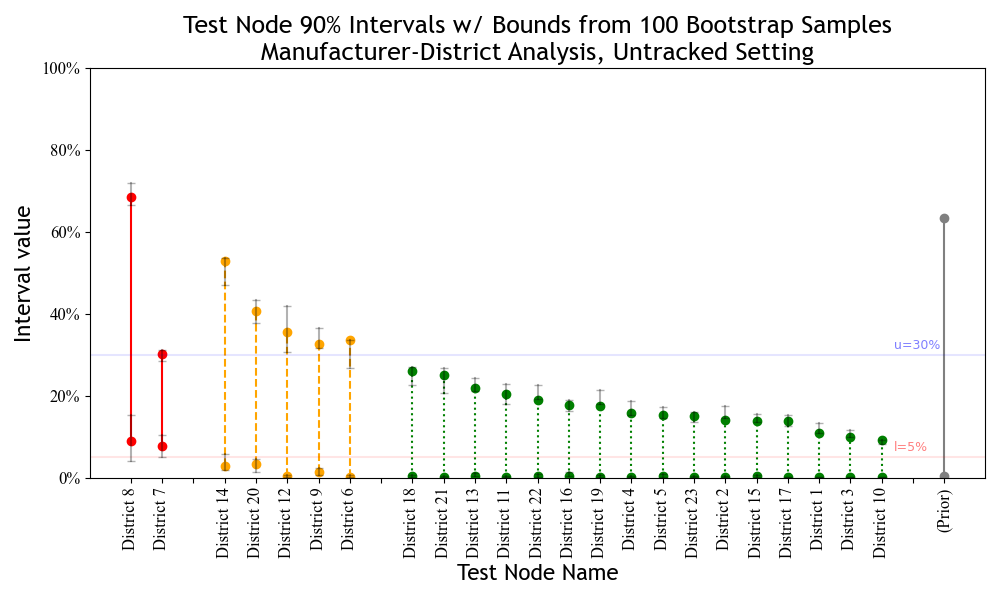}}
    \\
    \subfloat
    {\includegraphics[width=0.95\textwidth]{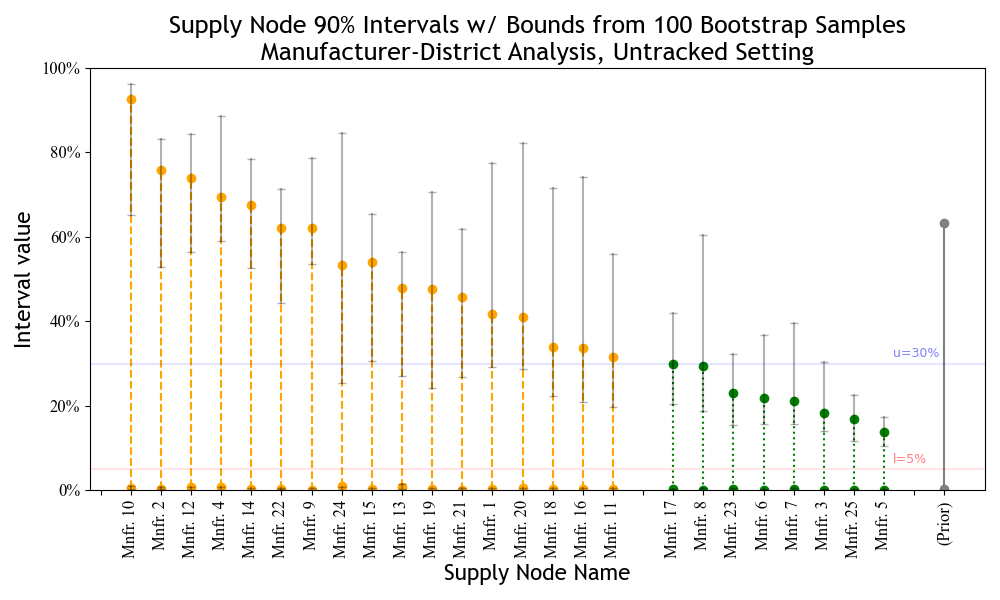} }
    \caption{Test-node 90\% intervals for MCMC samples generated using case-study data, including 90\% intervals for the upper and lower bounds for 100 bootstrap estimates of $\transMat$.
    Intervals with lower bounds above $l=5\%$ are featured in solid lines on the left, intervals with lower bounds below $l$ and upper bounds above $u=30\%$ are featured in dashed lines in the middle, and all other intervals are featured in dotted lines on the right. Small differences from Figures \ref{fig:manudist_TN} and \ref{fig:manudist_SN} stem from MCMC sampler variation.
    } \label{fig:manudist_boot}
\end{figure}

We explore sensitivity to the estimation of $\transMat$ through bootstrap sampling of the tests used to construct $\transMat$.
Sets of 406 bootstrap samples from the 406 observations provide different estimates for $\transMat$, with data $\dataSet=(\bm{\testResultpoint},\bm{\outletLabelpoint},\bm{\transMat})$ then used to generate posterior draws.
Thus positive rates at each test node remain constant, but $\transMat$ varies.
For 100 such sets of bootstrap samples, the 5\% and 95\% quantiles of the posterior draws for each node are stored.
The 5\% and 95\% quantiles of these quantiles across the 100 bootstrap sets are depicted as error bars in Figure \ref{fig:manudist_boot}.
The error bars for SFP intervals associated with test nodes change modestly with different $\transMat$ estimates: the upper and lower interval ends change by two or three percentage points.
District 8 and District 12 exhibit the largest sensitivity to $\transMat$ estimation, with the upper interval end for District 12 and the lower interval end for District 8 varying by more than ten percentage points.
The sensitivity of the intervals for these test nodes is likely due to their low prevalence in the data set: the data exhibit twelve tests associated with District 8 and only one test associated with District 12.
Considering the supply nodes, the error bars for SFP intervals associated with supply nodes are much wider than the error bars observed with test nodes.
The error bars for the upper interval end associated with Manufacturer 24 has a range of more than sixty percentage points.
The large fluctuations for inference of supply node SFP rates aligns with having partial information of the supply node associated with each test; analysis of untracked data thus carries inherent risk.
Effective untracked inference requires either favorable sourcing patterns by test nodes, as discussed in Section \ref{subsec:casestudyanalysis}, or more testing data.
Practitioners conducting untracked inference could consider using similar bootstrap techniques to examine the sensitivity of their analyses to the estimation of $\transMat$.

In the untracked analysis of the case study, 200,000 MCMC draws with 5,000 warm-start draws seem required to avoid any switching in the ranks of supply nodes, with the rank being where supply nodes align relative to each other on the horizontal axis.
The occurrence of rank switching aligns with the inference variance observed for supply nodes in the untracked case. 
For supply nodes in the tracked case, as well as test nodes generally, using 1,000 draws after 5,000 warm-start draws looks to produce the same rankings consistently for this case study.

\end{document}